\documentclass{article}

\usepackage{amssymb,amsfonts,amsmath}
\usepackage{cite,enumerate,float,indentfirst}
\usepackage{color}
\usepackage{booktabs}
    \newcommand{\tabitem}{~~\llap{\textbullet}~~}

\def\be{\begin{eqnarray}}
\def\ee{\end{eqnarray}}
\def\nn{\nonumber}

\def\tr{{\rm tr}\,}

\def\beq{\be }
\def\eeq{\ee}
\def\beqa{\be }
\def\eeqa{\ee}
\def\CR{\nn\\}

\def\Sch{{\rm Schur}}

\def\M{{\rm Md}}

\definecolor{red}{rgb}{1,0,0}
\definecolor{orange}{rgb}{1,0.5,0}
\definecolor{violet}{rgb}{0.7,0,1}

\def\bp{\bar{\bf p}}
\def\bp{\underline{\bf p}}

\def\LP{LP\ }

\hoffset= - 1.0in         

\begin{document}

\title{
\LARGE{ \bf Can tangle calculus be applicable to hyperpolynomials?
}}

\author{
{\bf Hidetoshi Awata$^a$}\footnote{awata@math.nagoya-u.ac.jp},
\ {\bf Hiroaki Kanno$^{a,b}$}\footnote{kanno@math.nagoya-u.ac.jp},
\ {\bf Andrei Mironov$^{c,d,e}$}\footnote{mironov@lpi.ru; mironov@itep.ru},
\ and \  {\bf Alexei Morozov$^{f,d,e}$}\thanks{morozov@itep.ru}
\date{ }
}

\maketitle

\vspace{-5.4cm}

\begin{center}
\hfill FIAN/TD-02/19\\
\hfill IITP/TH-05/19\\
\hfill ITEP/TH-09/19\\
\hfill MIPT/TH-05/19
\end{center}

\vspace{2.7cm}

\begin{center}
$^a$ {\small {\it Graduate School of Mathematics, Nagoya University,
Nagoya, 464-8602, Japan}}\\
$^b$ {\small {\it KMI, Nagoya University,
Nagoya, 464-8602, Japan}}\\
$^c$ {\small {\it Lebedev Physics Institute, Moscow 119991, Russia}}\\
$^d$ {\small {\it ITEP, Moscow 117218, Russia}}\\
$^e$ {\small {\it Institute for Information Transmission Problems, Moscow 127994, Russia}}\\
$^f$ {\small {\it MIPT, Dolgoprudny, 141701, Russia}}
\end{center}

\vspace{.5cm}

\begin{abstract}
We make a new attempt at the recently suggested program to express
knot polynomials through topological vertices, which can be considered
as a possible approach to the tangle calculus: we discuss the Macdonald deformation of
the relation between the convolution of two topological vertices and
the HOMFLY-PT invariant of the 4-component link $L_{8n8}$,
which both depend on four arbitrary representations.
The key point is that both of these are related to the Hopf polynomials in {\it composite} representations,
which are in turn expressed through composite Schur functions.
The latter are further
expressed through the skew Schur polynomials via the remarkable
Koike formula.
It is this decomposition that breaks under the Macdonald deformation
and gets restored only in the (large $N$) limit of $A^{\pm 1}\longrightarrow 0$.
Another problem is that the Hopf polynomials in the composite representations in the refined case are ``chiral bilinears''
of Macdonald polynomials, while convolutions of topological vertices
involve ``non-chiral combinations'' with one of the Macdonald polynomials entering with permuted $t$ and $q$.
There are also other mismatches between the Hopf polynomials in the composite representation and the topological 4-point function in the refined case, which we discuss.
\end{abstract}

\section{Introduction}

Recently in \cite{L8n8} we started a new program
to construct link polynomials (\LP\!\!)\footnote{
HOMFLY-PT invariants and their $q,t$-deformations
with the help of Macdonald functions (``hyperpolynomials'')
for {\it links} are not quite {\it polynomials},
which is also not the case for {\it un}reduced invariants of knots,
this is because of easily controlled  factors in denominators
expressed through dimensions, quantum or Macdonald, therefore calling them
``polynomials'' is a rather light abuse of terminology, which is now broadly accepted.
}
\cite{knotpols} from topological vertices  (TV) \cite{topvert}.
It is one of possible realizations of the tangle calculus \cite{tangles},
and it is also important for a still missed reformulation of knot calculus
in the standard language of matrix models \cite{BEMS}.
As the first example, we took in \cite{L8n8} the 4-component link $L_{8n8}$,
because \LP in this case are essentially made of those for the Hopf link,
when they are just convolutions of two TV.\footnote{
In fact the well known \LP for the Hopf link were actually used to {\it deduce}
explicit expressions for the TV, though nowadays there are more direct ways to do it
directly from representation theory of the DIM algebra \cite{DIM,Miki,AFS}.
}
Though conceptually the relation between the \LP and the TV is very plausible and long expected,
a practical comparison is somewhat difficult.
On of the main reasons is that related to the TV are the {\it colored} \LP\!\!,
moreover, the relevant ones are those in {\it composite} representations,
which essentially depend on the rank of the gauge algebra, i.e. on $N$ for $SL_N$.
In such situations, the very {\it definition} of $A$-dependent \LP ($A=q^N$ or $t^N$)
requires some care and involves a kind of projective limit,
which leads to what we call the {\it uniform} \LP, see \cite{uni}.
Therefore before going deep into the study of TV $\longrightarrow$ \LP constructions,
one should better understand how this {\it uniformization} works.
In the case of the HOMFLY-PT \LP for the Hopf link,
it was thoroughly investigated in \cite{MMhopf},
and this was what made possible the $L_{8n8}$ study in \cite{L8n8}.
The natural next step is to lift the construction from the HOMFLY-PT  to
{\it hyper}polynomials, i.e. from the Schur polynomial based expressions to
those based on the Macdonald polynomials and thus depending on three rather than two
parameters: $A,\ q$ and $t$.

Unfortunately this runs into a set of problems,
which we discuss in the present paper.

The {\bf first problem} is the recently demonstrated \cite{compomac}
complexity of {\it uniformization} for the composite Macdonald polynomials\footnote{
Studied in \cite{compomac} were actually the much more general Kerov functions
in composite representations, which helped to demonstrate
what a non-trivial property {\it uniformization} is.
It is absent for the generic Kerov functions (the $N$-dependence can not be efficiently
encoded in terms of a controllable dependence on a parameter like $A$)
and emerges as a miracle only at the Macdonald locus.
}
and a failure of the Koike formula \cite{Koike,Kanno,MMhopf},
which describes the composite Schur functions
and plays the central role in the HOMFLY-PT case.
We review the Schur case and the problem
with Macdonald lifting in sec.\ref{compMac}.
Deviations from the Koike formula are two-fold:
there are many more terms in decomposition of the Macdonald composites
and the coefficients contain new $A$-dependent denominators.
The first issue is not obligatory a disaster, because even in the Schur case
the convolution of TV provided a sum over composite Hopf LP\!,
so that the $L_{8n8}$ LP for given representations was just a single
(``dominant") term in this expansion, thus one could hope that
this ``projection" just gets somewhat more involved after the $q,t$-deformation.
However, the $A$-dependent denominators are far less pleasant:
they have few chances to be reproduced within the TV calculus.
The story is actually quite a usual one for {\it super}polynomials,
where answers at particular $N$ (Khovanov-Rozansky polynomials \cite{KhR})
are related to the $A$-dependent quantities by a conceptually obscure
DGR trick \cite{DGR}.
It looks like, perhaps unexpectedly, the situation is going to be the same even
for the {\it hyper}polynomials, at least  in the {\it composite} (i.e. $N$-dependent)
representations.

The {\bf second problem} is a peculiar chirality of the Macdonald polynomials,
which are not symmetric under permuting the $q$ and $t$ parameters:
$\overline{M}_{q,t}\{p\}:=M_{t,q}\{p\} \neq M_{q,t}\{p\}$.
As we explain in sec.\ref{s6}, the Hopf LP are chiral bilinears,
$M\cdot M$, while the convolutions of TV are non-chiral $M\cdot\overline{M}$, and thus they are different.
This is not an innocent problem: non-chirality in convolutions is
required by a counterpart of the charge conservation: the 2-point functions
$\langle e^{i\alpha\phi(x)}\ e^{i\beta\phi(0)}\rangle \neq 0$ for $\beta=-\alpha$.
Thus building a chiral bilinear as an average in the network models \cite{network,AKM4OZ}
is a quite a challenge requiring new ideas and insights.

Last, but not least, another recent development \cite{KNTZ}
was a demonstration that rectangularly colored knot hyperpolynomials
in the particular case of twist knots can be related to
the {\it shifted} rather than {\it skew} Macdonald polynomials.
There is no true difference between those in the Schur (HOMFLY-PT) case,
but there is after the $q,t$-deformation.
One can not exclude that this change $skew\longrightarrow shifted$
is the right thing to do not only for the twisted knots and not only for rectangular
representations, and one should look for decompositions of composites
and for the refinement of TV in these terms.
This, however, remains a subject for the future work.

\paragraph{Notations.}  Various quantities introduced and used throughout the text are:

\bigskip

\begin{tabular}{ll}
& \underline{Time variables:}\\
&\\
    \tabitem $h_\mu$:\ \ \  the Macdonald hook length, (\ref{hook}) & \tabitem $p_k^{(\mu)}$:\ \ \ (\ref{pmu}) \\
[.5\normalbaselineskip]
    \tabitem $f_\mu$:\ \ \     the framing factor due to Taki \cite{Taki}, (\ref{fr}) & \tabitem ${\bf p}^{*(\mu,\lambda)}_k$:
     \ \ \ (\ref{pU1})\\
[.5\normalbaselineskip]
     \tabitem $||\mu||$:\ \ \ the norm of the Macdonald polynomial $M_\mu$, (\ref{MOR})    &  \tabitem ${\bp}^{*(\mu,\lambda)}_k:={\bf p}^{*(\mu,\lambda)}_k(A^{-1},q^{-1},t^{-1})$ \\
[.5\normalbaselineskip]
\tabitem $D_{(\mu,\lambda)}$:\ \ \    a factorized combination, (\ref{65})    & \underline{Topological 4-point functions:}
\\
[.5\normalbaselineskip]
\tabitem $\M_\mu$:\ \ \    the Macdonald dimension, (\ref{nu})    & \tabitem ${\bf Z}\left[\begin{array}{c|c}\!\! \mu_1& \lambda_2\\
\lambda_1& \mu_2\end{array}\right]$:\ \ \  (\ref{4p}) and (\ref{Z})\\
[.5\normalbaselineskip]
\tabitem $\Sigma:=|\mu_1|+|\mu_2|+|\lambda_1|+|\lambda_2|$&\tabitem ${\bf Z'}\left[\begin{array}{c|c}\mu_1\,\lambda_1&\!\! \\\!\! &\mu_2\,\lambda_2\end{array}\right]$:\ \ \ (\ref{4plambda}) and (\ref{Zp})\\
[.5\normalbaselineskip]
\tabitem ${\cal P}_{\mu,\lambda}^{\rm Hopf}$:\ \ \ the Hopf hyperpolynomial, (\ref{Hopf}) and (\ref{RJ})&
\tabitem ${\bf Z''}\left[\begin{array}{c|c}\!\! \lambda_3 & \lambda_4 \\
\lambda_1& \lambda_2\end{array}\right]$:\ \ \ (\ref{4pmu})\\
[.5\normalbaselineskip]
\tabitem $\left[{\cal P}_{\mu,\lambda}^{\rm Hopf}\right]_0$:\ \ \ the Hopf invariant leading behaviour at $A\to 0$ &
\tabitem $\widehat{\bf Z}\left[\begin{array}{c|c}\!\! \bullet& \bullet\\
\bullet& \bullet\end{array}\right]:={\bf Z}\left[\begin{array}{c|c}\!\! \bullet& \bullet\\
\bullet& \bullet\end{array}\right]\cdot{\bf Z}\left[\begin{array}{c|c}\!\! \varnothing& \varnothing\\
\varnothing& \varnothing\end{array}\right]^{-1}$\\
[\normalbaselineskip]
& \phantom{SH}and similarly for ${\bf Z'}$ and ${\bf Z''}$\\
  \end{tabular}

We also use the following notation:
\be
\{x\}:=&x-x^{-1}\nn\\
\overline{X}(q,t):=&X(t,q)\nn\\
F(x)&\ \ \ \ \ \ \ \hbox{a symmetric function of variables }x_i\nn\\
F\{p_k\}&\ \ \ \ \ \ \hbox{a symmetric function as a function of the power sums }p_k\nn\\
R^\vee&\ \ \ \ \ \ \ \hbox{the Young diagram conjugate to }R \nn\\
|R|&\ \ \ \ \ \ \ \hbox{size of the Young diagram} \nn
\ee
We choose the standard order of arguments $(A,q,t)$. This means that, e.g., the notation $X(A,t,q)$ denotes the permuted arguments $q$ and $t$.
Various quantities related with the Macdonald polynomials can be found in the Appendix. Throughout the text, we omit the $U(1)$-factor from the normalization of the Hopf hyperpolynomial. We also sometimes omit the subscript $k$ of time variables $p_k$, when it should not lead to a misunderstanding.

Notice that we make a change of parameters $(q,t)\to (q^2,t^2)$ in the Macdonald polynomials as compared with \cite{Mac}.

\bigskip

\section{The outline of the paper}

\subsection{Summary of \cite{L8n8}}

In \cite{L8n8}, we studied the interplay of three objects:
\begin{itemize}
\item[\bf A.] The four-point function,
\be
Z_{\mu_1,\mu_2;\lambda_1,\lambda_2} = \sum_\xi (-A^2)^{|\xi|} C_{\xi\mu_1\lambda_1} C_{\xi^\vee\mu_2\lambda_2}
\ee
\begin{picture}(200,0)(-400,-10)
\put(0,0){\line(1,1){30}}
\put(30,30){\line(1,0){30}}
\put(30,30){\line(0,1){30}}
\put(0,0){\line(-1,0){30}}
\put(0,0){\line(0,-1){30}}
\put(10,10){\line(-1,-1){2}}
\put(40,30){\line(-1,0){2}}
\put(30,40){\line(0,-1){2}}
\put(-10,0){\line(1,0){2}}
\put(0,-10){\line(0,1){2}}
\put(10,20){\mbox{$\xi$}}
\put(55,35){\mbox{$\lambda_1$}}
\put(15,55){\mbox{$\mu_1$}}
\put(-28,5){\mbox{$\lambda_2$}}
\put(5,-28){\mbox{$\mu_2$}}
\end{picture}

\noindent
obtained by the convolution of two topological vertices,
\be
C_{\xi\mu\lambda} =
q^{\nu'(\lambda)-\nu(\lambda)}\cdot \Sch_{\mu}\{p^{(\varnothing)}\}\cdot
\sum_{\eta} \Sch_{\xi/\eta} \{p^{(\mu)}\}\cdot  \Sch_{\lambda^\vee/\eta} \{p^{(\mu^\vee)}\}
\ee
 (for the notation, see (\ref{nu}) and (\ref{pmu}) at $t=q$).
\item[\bf B.] The HOMFLY-PT polynomial for the four-component link $L_{8n8}$, which depends on four
representations and coincides with the HOMFLY-PT polynomial of the 2-component Hopf link:
\be
{\cal H}^{L_{8n8}}_{\mu_1,\lambda_1,\mu_2,\lambda_2} =
\sum_{\stackrel{\lambda\in \lambda_1\otimes\overline{\lambda}_2}{\mu\in \mu_1\otimes\overline{\mu}_2}} \ \
N^{\lambda}_{\lambda_1\lambda_2}\cdot N^\mu_{\mu_1\mu_2}\cdot
{\cal H}^{\rm Hopf}_{\lambda,\mu}
\label{L8n8}
\ee
\parbox{7cm}{as follows from the picture,
where $L_{8n8}$ is on the left and Hopf = ${\rm Torus}_{[2,2]}$ is on the right ($N^\mu_{\mu_1\mu_2}$ are the Littlewood-Richardson coefficients):}

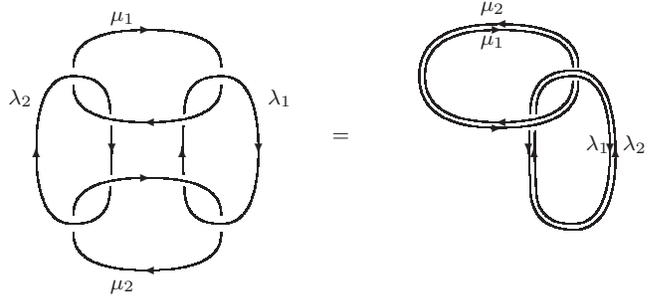
\begin{picture}(300,70)(-280,-100)
\unitlength=.7pt
\qbezier(-40,0)(-40,20)(0,20) \qbezier(40,0)(40,20)(0,20)
\qbezier(-40,-10)(-40,-30)(0,-30) \qbezier(40,-10)(40,-30)(0,-30)

\put(0,-80){
\qbezier(-40,0)(-40,20)(0,20) \qbezier(40,0)(40,20)(0,20)
\qbezier(-40,-10)(-40,-30)(0,-30) \qbezier(40,-10)(40,-30)(0,-30)
}

\qbezier(-40,-5)(-60,-5)(-60,-45)\qbezier(-40,-85)(-60,-85)(-60,-45)
\qbezier(-40,-5)(-20,-5)(-20,-25)\qbezier(-40,-85)(-20,-85)(-20,-65)
\qbezier(-19.5,-32)(-19,-45)(-19.5,-58)

\qbezier(40,-5)(60,-5)(60,-45)\qbezier(40,-85)(60,-85)(60,-45)
\qbezier(40,-5)(20,-5)(20,-25)\qbezier(40,-85)(20,-85)(20,-65)
\qbezier(19.5,-32)(19,-45)(19.5,-58)

\put(-60,-45){\vector(0,1){2}}
\put(-19,-45){\vector(0,-1){2}}
\put(0,20){\vector(1,0){2}}
\put(0,-30){\vector(-1,0){2}}

\put(60,-45){\vector(0,-1){2}}
\put(19,-45){\vector(0,1){2}}
\put(0,-110){\vector(-1,0){2}}
\put(0,-60){\vector(1,0){2}}
\footnotesize{
\put(-75,-20){\mbox{$\lambda_2$}}
\put(-20,25){\mbox{$\mu_1$}}
\put(65,-20){\mbox{$\lambda_1$}}
\put(-20,-120){\mbox{$\mu_2$}}
}

\put(100,-40){\mbox{$=$}}

\put(190,0){
\qbezier(-40,-5)(-40,20)(0,20) \qbezier(40,0)(40,20)(0,20)
\qbezier(-40,-5)(-40,-30)(0,-30) \qbezier(40,-8)(40,-30)(0,-30)

\qbezier(-43,-5)(-43,23)(0,23) \qbezier(43,0)(43,23)(0,23)
\qbezier(-43,-5)(-43,-33)(0,-33) \qbezier(43,-8)(43,-33)(0,-33)

\qbezier(40,-5)(60,-5)(60,-45)\qbezier(40,-85)(60,-85)(60,-45)
\qbezier(40,-5)(20,-5)(20,-25)\qbezier(40,-85)(20,-85)(20,-65)
\qbezier(19.5,-35)(19,-45)(20,-65)

\qbezier(43,-2)(63,-5)(63,-48)\qbezier(43,-88)(63,-85)(63,-48)
\qbezier(43,-2)(17,-1)(17,-25)\qbezier(43,-88)(17,-89)(17,-65)
\qbezier(19.5,-35)(19,-45)(20,-65)
\qbezier(16.5,-35)(16,-45)(17,-65)

\put(0,20){\vector(1,0){2}}
\put(0,-30){\vector(-1,0){2}}

\put(0,23){\vector(-1,0){2}}
\put(0,-33){\vector(1,0){2}}

\put(60,-45){\vector(0,-1){2}}
\put(19,-45){\vector(0,1){2}}

\put(63,-45){\vector(0,1){2}}
\put(16,-45){\vector(0,-1){2}}

\footnotesize{
\put(47,-45){\mbox{$\lambda_1$}}
\put(-10,30){\mbox{$\mu_2$}}
\put(67,-45){\mbox{$\lambda_2$}}
\put(-10,12){\mbox{$\mu_1$}}
}
}

\end{picture}
\item[\bf C.] The HOMFLY-PT polynomial for the Hopf link in composite representations (see (\ref{p*comp}) at $t=q$):
\be
 {\cal H}^{\rm Hopf}_{(\mu_1,\mu_2)\times(\lambda_1,\lambda_2)}
= {\rm Qd}_{(\mu_1,\mu_2)}\cdot {\rm Schur}_{(\lambda_1,\lambda_2)}\{{\bf p}^{*(\mu_1,\mu_2)}\}
= {\rm Qd}_{(\lambda_1,\lambda_2)}\cdot {\rm Schur}_{(\mu_1,\mu_2)}\{{\bf p}^{*(\lambda_1,\lambda_2)}\}
\label{comHopf}
\ee
which involves a peculiar Schur function in the composite representation $(\mu_1,\mu_2)$ (see (\ref{pV}) at $t=q$)
\be
{\rm Schur}_{(\mu_1,\mu_2)}\{{\bf p}^{*V}\}
= \sum_\eta (-)^{|\eta|}\cdot {\rm Schur}_{\mu_1/\eta}\{{\bf p}^{*V}\}\cdot
{\rm Schur}_{\mu_2/\eta^\vee}\{\bp^{*V}\}
\label{comSchur}
\ee
Here ${\rm Qd}_{(\mu_1,\mu_2)}$ are the quantum dimensions of the composite representation $(\mu_1,\mu_2)$ \cite{MMhopf}. We call this formula the Koike formula, \cite{Koike}.
\end{itemize}
The {\it central claim} of \cite{L8n8} was that {\bf the four-point function coincides
with the {\it dominant} contribution to the sum in ${\cal H}^{L_{8n8}}$,
which is exactly the Hopf polynomial colored with two composite representations,
${\cal H}^{\rm Hopf}_{(\mu_1,\mu_2)\times(\lambda_1,\lambda_2)}$},
\be\label{eq}
\boxed{
{\cal G}^{L_{8n8}}_{\mu_1\times\lambda_1\times\mu_2\times\lambda_2}: =
\hbox{\bf Pr}\Big[\underbrace{{\cal H}^{L_{8n8}}_{\mu_1,\lambda_1,\mu_2,\lambda_2}
}_{L_{8n8}\ \text{   polynomial}}
\Big]_{\rm max}=
\frac{\overbrace{Z_{\mu_1,\mu_2;\lambda_1,\lambda_2}}^{\text{4-point function}}}
{Z_{\varnothing,\varnothing:\varnothing, \varnothing}}
=\overbrace{{\cal H}^{\rm Hopf}_{(\lambda_1,\lambda_2)\times(\mu_1,\mu_2)}
}^{\text{ Hopf polynomial}}
}
\ee

\subsection{Macdonald deformation}

The goal of the present paper is to consider a modification of all these ingredients
under the $(q,t)$-deformation which changes the Schur to the Macdonald polynomials,
and the HOMFLY-PT polynomials to the {\it hyper}polynomials \cite{AS,DMMSS,Che}.
\begin{itemize}
\item[\bf A$'$.] The Macdonald deformation(s) of $Z$ is well known: it is made from
the convolution of refined topological vertices \cite{rtv,AK08}.
However, there are many problems beginning with existence of different
functions that are not obviously equivalent,
and ending with asymmetry of the emerging refined 4-point function.
\item[\bf B$'$.] The topological relation (\ref{L8n8}) between the $L_{8n8}$ and Hopf links now
has no any {\it immediate} implication for the super- and hyper-polynomials:
it is yet unclear if the former exist at all in arbitrary
(non-antisymmetric) representations,
and if the latter can be defined for arbitrary knots and links and
are invariants under the Reidemeister moves.
\item[\bf C$'$.] Expression like (\ref{comHopf}) exists for the Hopf hyperpolynomials
in ordinary representations, but its lifting to composite representations
is a non-trivial issue.
Whatever it is, it is {\it inconsistent} with (\ref{comSchur}):
the composite Hopf hyperpolynomial is {\it not} decomposed into skew Macdonald functions.
\end{itemize}

\subsection{Macdonald deformation breaks (\ref{eq})}

It turns out that, in the refined case, the correspondence (\ref{eq}) is no longer correct. In fact, there are several problems related to this case. We list them here, and discuss the details further in this paper.
\begin{itemize}
\item[{\bf 1.}] First of all, in the refined case, there are several distinct 4-point functions that have the same unrefined $t=q$ limit. Indeed, one can produce the 4-point function convolving two topological vertices (\ref{rtv}) in the third, second, or first index, with the second index corresponding to the preferred direction \cite{AK08,AFS}, they all coincide in the unrefined case due to the cyclic symmetry of the topological vertex in that case. We denote these three possibilities ${\bf Z}$, ${\bf Z''}$ and ${\bf Z'}$  accordingly (see  (\ref{4p}),  (\ref{4pmu}) and (\ref{4plambda})). The rules of constructing the 4-point functions can be directly obtained from the quantum toroidal (Ding-Iohara-Miki) algebra description, with the topological vertex being an intertwining operator of three Fock representations \cite{AFS}.

    In the refined case, these three functions are different. However, in the case of two non-trivial and two trivial representations, different 4-point functions still describe the (refined) Hopf link invariant, the Hopf hyperpolynomial \cite{AS,DMMSS,Che,IK}. For instance,
\be\label{7}
\widehat{\bf Z}\left[\begin{array}{c|c}\!\! \mu& \lambda\\
\varnothing& \varnothing\end{array}\right](A,q,t)=
\widehat{\bf Z}\left[\begin{array}{c|c}\!\! \varnothing& \varnothing\\
\lambda& \mu\end{array}\right]
(Aq/t,t,q)=
(-1)^{|\lambda|}A^{|\mu|+|\lambda|}f_\mu\cdot C_\mu(A^{-1},t,q)\cdot
{\cal P}^{\rm Hopf}_{\mu,\lambda^\vee}(A^{-1},q^{-1},t^{-1})
\ee
where the invariant on the r.h.s. describes the mirror-reflected Hopf link. Similarly, for the other 4-point function,
\be\label{8}
\widehat{\bf Z'}\left[\begin{array}{c|c}\varnothing\,\varnothing&\!\! \\\!\! &\mu\,\lambda\end{array}\right](A,q,t)=
\widehat{\bf Z'}\left[\begin{array}{c|c}\mu\,\lambda&\!\! \\\!\! &\varnothing\,\varnothing\end{array}\right](Aq/t,t,q)=
\left(A{q\over t}\right)^{|\mu|+|\lambda|}f_{\lambda^\vee}\cdot C_\mu(A,q,t)\cdot
{\cal P}^{\rm Hopf}_{\mu,\lambda}(A,t^{-1},q^{-1})
\ee
and for the third 4-point function (with the internal edge along the preferred direction)  {\bf this is correct only at $\lambda=[1]$}:
\be\label{9}
\widehat{\bf Z''}\left[\begin{array}{c|c}\!\! \varnothing & \varnothing \\
\lambda& \mu\end{array}\right](A,q,t)=\widehat{\bf Z''}\left[\begin{array}{c|c}\!\! \mu & \lambda \\
\varnothing& \varnothing\end{array}\right](Aq/t,t,q)
\stackrel{\lambda=[1]}{=}\left(-{q\over t}\right)^{2|\mu|+1}A^{|\mu|+1}\cdot {\cal P}^{\rm Hopf}_{\mu,[1]}(A^{-1},t^{-1},q^{-1})
\ee
The latter invariant on the r.h.s. again describes the mirror-reflected Hopf link.

The coefficients are the ratios of products of simple factors,
\be
C_\mu(A,q,t):={\M_{\mu^\vee}(Aq/t,q,t)\over\M_{\mu^\vee}(A,q,t)}
\ee

\item[{\bf 2.}]
However, different 4-point functions {\bf far do not coincide}
when all the 4 representations are non-trivial, the case expected to correspond to the Hopf hyperpolynomial in the {\it composite} representation. Moreover, the difference is quite strong and manifests itself in different ways.

\item[{\bf 3.}]  Coincidence disappears already in the case of
two non-trivial representations in the both preferred directions or both unpreferred directions (see sec.5.3), which describes, in the unrefined case, the quantum dimension of the composite representation:
\be
{\bf Z}\left[\begin{array}{c|c}\!\! \mu_1& \varnothing\\
\varnothing& \mu_2\end{array}\right]
;\ {\bf Z}\left[\begin{array}{c|c}\!\! \varnothing& \mu_2\\
\mu_1& \varnothing\end{array}\right]
\ne
{\cal P}^{\rm Hopf}_{(\mu_1,\mu_2),(\varnothing,\varnothing)}
\ee
where the last quantity is still equal to the Macdonald dimension of the composite representation $(\mu_1,\mu_2)$.

\item[{\bf 4.}] Another drawback of the refined 4-point functions is that they are not symmetric in the case of four different representations under the permutation $(\mu_1,\mu_2)\leftrightarrow (\lambda_1,\lambda_2)$, which is the case for the Hopf link invariant, since it corresponds to just interchanging the Hopf link components.  Instead of this, there is only the property
\be\label{12a}
\widehat{\bf Z}\left[\begin{array}{c|c}\!\! \mu_1& \lambda_2\\
\lambda_1& \mu_2\end{array}\right](A,q,t)=
\widehat{\bf Z}\left[\begin{array}{c|c}\!\! \mu_2& \lambda_1\\
\lambda_2& \mu_1\end{array}\right](Aq/t,t,q)
\ee

\item[{\bf 5.}]
Discrepancy between the full-fledged 4-point functions and Hopf polynomial,
which is drastic for generic $A$, nearly disappears in the limits when
$A\to 0$ or $A\to \infty$.
Still an important difference persists.
The refined topological string functions
have a ``chirality'' structure different from that of the Hopf hyperpolynomial:
both are decomposed, but differently.
The hyperpolynomial becomes
\be
{\cal P}_{(\mu_1,\mu_2),(\lambda_1,\lambda_2)}^{\rm Hopf}\ \stackrel{A\to 0}{\sim}
\left[{\cal P}_{\lambda_1,\mu_1}^{\rm Hopf}\right]_0\cdot
\left[{\cal P}_{\lambda_2,\mu_2}^{\rm Hopf}\right]_0
\ee
where we denoted the Hopf hyperpolynomial at $A\to 0$ as
$\left[{\cal P}_{\lambda_2,\mu_2}^{\rm Hopf}\right]_0$, while the 4-point function is
\be
{\bf Z'}\left[\begin{array}{c|c}\mu_1\,\lambda_1&\!\! \\\!\! &\mu_2\,\lambda_2\end{array}\right]\stackrel{A\to 0}{\sim}
\left[\,\overline{{\cal P}_{\lambda_1^\vee,\mu_1^\vee}^{\rm Hopf}}\,\right]_0\cdot
\left[{\cal P}_{\lambda_2^\vee,\mu_2^\vee}^{\rm Hopf}\right]_0
\ee
Here bar denotes a peculiar conjugation, provided by the permutation of $q$ and $t$.
One can see that this string function is a {\it non-chiral} product
of the two factors with permuted $q$ and $t$ with respect to each other.
At the same time, these two factors are of the same {\it chirality} in the Hopf case.
\end{itemize}

\noindent
In the next subsection \ref{firlev}, we provide the simplest example to illustrate these problems
and explain why they are not so easy to avoid.

\subsection{First level contributions to the partition function \label{firlev}}

One may think that the problems just described could be cured by changing the basis: in principle, one has to compare \cite{IK} the generating function of Hopf link invariants in the refined Chern-Simons theory,
\be
Z_{CS}(U_1,U_2,U_3,U_4):=\sum_{\mu_i,\lambda_i}{\cal P}^{\rm Hopf}_{(\mu_1,\mu_2),(\lambda_1,\lambda_2)}\overline{M}_{\mu_1}(U_1)
\overline{M}_{\mu_2}(U_2)\overline{M}_{\lambda_1}(U_3)\overline{M}_{\lambda_2}(U_4)
\ee
and the open string partition function with four holonomies $U_i$, $i=1,2,3,4$,
\be
Z_{str}(U_1,U_2,U_3,U_4):=\sum_{\mu_i,\lambda_i}
\widehat{\bf Z}\left[\begin{array}{c|c}\!\! \mu_1& \lambda_2\\
\lambda_1& \mu_2\end{array}\right]\cdot V_{\mu_1}(U_1)
V_{\mu_2}(U_2) V_{\lambda_1}(U_3)V_{\lambda_2}(U_4)
\ee
where $V_\mu(U)$ is a (graded) basis of symmetric functions. Item 5 of the previous subsection makes a hint that some of these functions would be better not $\overline{M}_{\mu}(U)$, but substituted by $M_{\mu}(U)$, or $M_{\mu^\vee}(U)$ in order to compare with the Hopf polynomial. We discuss this issue in sec.\ref{lit}.

However, it does not solve the problem: these two partition functions do not coincide. Indeed, consider the first level representations, where there is only one symmetric function and, hence, the corresponding coefficients of the partition functions should be proportional to each other.  This is, however, not the case. Indeed, when one of the Hopf components is colored with the fundamental representation, and the other one with the adjoint representation (see the notations in (\ref{def}), (\ref{Dadj})),
\be
{\cal P}_{[1],([1],[1])}={\M_{([1],[1])}\over \{t\}}\cdot\left[\Big(\underline{q^2-(q/t)^2+t^{-2}}\Big)A-\Big(\underline{t^2-(t/q)^2+q^{-2}}\Big)A^{-1}\right]
\ee
while
\be\label{18}
\widehat{\bf Z}\left[\begin{array}{c|c}\!\! [1]& \varnothing\cr
[1]& [1]\end{array}\right]\sim D_{([1],[1])}\left[\Big(\underline{t^2-(t/q)^2+q^{-2}}\Big)
{\{Aq/t\}\over \{q\}}\right]
\ee

These two expressions are different!
Moreover,
\be\label{19}
\widehat{\bf Z}\left[\begin{array}{c|c}\!\! [1]&[1]\cr
[1]& \varnothing\end{array}\right] \sim {\{Aq/t\}\over\{t\}^2\{q\}}
\left[\left(qA \right)^2  \big(\underline{q^2 -(q/t)^2+t^{-2}}\big)
-(1+q^2t^2 )\big(\underline{ t^2-(t/q)^2+q^{-2}}\big)
+\left(A/t\right)^{-2}  \big(\underline{t^2-(t/q)^2+ q^{-2}}\big)\right]
\ee
and there are two more expressions with two other locations of $\varnothing$,
related to (\ref{18}) and (\ref{19}) by the rule (\ref{12a}).
This is just a manifestation of the fact that, while the Hopf hyperpolynomial is symmetric w.r.t. permuting the components, the four-point function is not! Hence, they are, indeed, different quantities.

Similarly, when the both representations are adjoint, the hyperpolynomial
\be
{\cal P}_{([1],[1]),([1],[1])}=D_{([1],[1])}R_{([1],[1])}^2\left(-1+{\{Aq/t\}\over
\{A\}\{q\}\{t\}}\left[\Big(qt+(qt)^{-1}\Big)\{Aq/t\}-\{Aq^2/t^2\}\right]^2\right)
=\nn \\
 =D_{([1],[1])}R_{([1],[1])}^2\left(-1+{\{Aq/t\}\over
\{A\}\{q\}\{t\}}\left[A\cdot\Big(\underline{q^2-(q/t)^2+t^{-2}}\Big)
-\frac{1}{A}\cdot \Big(\underline{t^2-(t/q)^2+q^{-2}}\Big)\right]^2\right)
\label{hypadjadj}
\ee
while the 4-point function is
\be
\widehat{\bf Z}\left[\begin{array}{c|c}\!\! [1]&[1]\cr
[1]&[1]\end{array}\right]\sim
D_{([1],[1])}\left(-1+\left[\Big(q^{-2}-(q/t)^2+t^2\Big)
{\{Aq/t\}\over \{q\}}\right]\cdot \left[\Big(q^2-(t/q)^2+t^{-2}\Big){\{A\}\over \{t\}} \right]\right)=\nn\\
=D_{([1],[1])}\left(-1+{\{A\}\{Aq/t\}\over \{q\}\{t\}} \Big(\underline{q^{-2}-(q/t)^2+t^2}\Big)\cdot\Big(\underline{q^2-(t/q)^2+t^{-2}}\Big)\right)
\label{4padjadj}
\ee
The two expressions are again different: they are constructed from the same elementary block $\xi:=q^2-(t/q)^2+t^{-2}$ and $\overline{\xi}$, but in different ways.
Surviving in the limits $A^{\pm 1} \to 0$ is the square of one of the two underlined
structures in (\ref{hypadjadj}), what makes the limits ``chiral",
while (\ref{4padjadj}) is rather a ``non-chiral'' product of two conjugate factors
(i.e. ``a modulus squared").

Similar formulas emerge for the other topological functions. For instance,
\be\label{3Z}
\widehat{\bf Z'}\left[\begin{array}{c|c}[1]\,[1]&\!\! \\\!\! &[1]\,[1]\end{array}\right]=\widehat{\bf Z''}\left[\begin{array}{c|c}\!\! [1] & [1] \cr
[1]& [1]\end{array}\right]=\widehat{\bf Z}\left[\begin{array}{c|c}\!\! [1]&[1]\cr
[1]&[1]\end{array}\right]
\ee
For the origin of these identities, see sec.\ref{lit}.

The structure of answers in this case is
\be
\widehat{\bf Z}\left[\begin{array}{c|c}\!\! [1]&[1]\cr
[1]&[1]\end{array}\right]\sim D_{([1],[1])}\cdot\Big(-1+\mathfrak{p}_1\mathfrak{p}'_1\Big)\\
{\cal P}^{\rm Hopf}_{adj,adj}\sim \M{([1],[1])}\cdot\Big(-{\{A\}\{q\}\over \{t\}\{Aq/t\}}+\mathsf{p}_1\mathsf{p}'_1\Big)
\ee
The difference is in three points: the common factor in the second formula is the Macdonald dimension of the adjoint representation, (\ref{def}), which is not equal to $D_{([1],[1])}$, (\ref{Dadj}); $-1$ in the first formula is substituted with a non-unit term in the second one, and, in the product of the first Macdonald polynomials, that is, of $p_1$, the time-variables are different.

In the next sections, we demonstrate that these problems remain  just the same for
generic quadruples of representations (i.e. pairs of composites),
no new complications seem to emerge;
thus their resolution, if any, at the level of two adjoint representations can be sufficient to understand
the situation in general.

\section{Composite Macdonald polynomials}

\subsection{Composite representations\label{comp}}

Composite are representation of $SL_N$ described by the $N$-dependent
Young diagram
$$(R,P)= \Big[r_1+p_1,\ldots,r_{l_R}+p_1,\underbrace{p_1,\ldots,p_1}_{N-l_{\!_R}-l_{\!_P}},
p_1-p_{_{l_{\!_P}}},p_1-p_{{l_{\!_P}-1}},\ldots,p_1-p_2\Big]
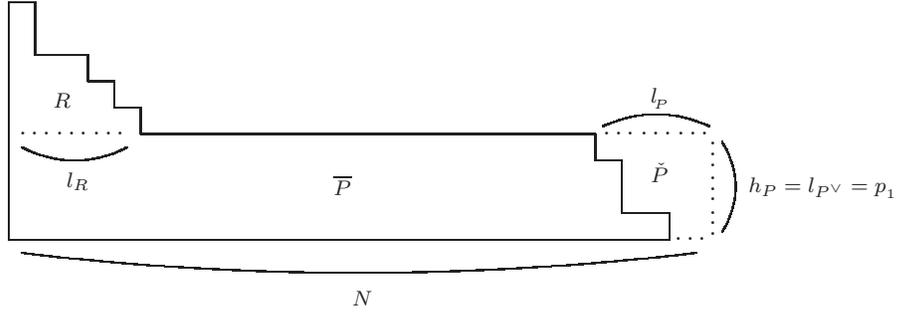
$$

\begin{picture}(300,125)(-90,-30)

\put(0,0){\line(0,1){90}}
\put(0,0){\line(1,0){250}}
\put(50,40){\line(1,0){172}}

\put(0,90){\line(1,0){10}}
\put(10,90){\line(0,-1){20}}
\put(10,70){\line(1,0){20}}
\put(30,70){\line(0,-1){10}}
\put(30,60){\line(1,0){10}}
\put(40,60){\line(0,-1){10}}
\put(40,50){\line(1,0){10}}
\put(50,50){\line(0,-1){10}}

\put(265,2){\mbox{$\vdots$}}
\put(265,15){\mbox{$\vdots$}}
\put(265,28){\mbox{$\vdots$}}

\put(252,0){\mbox{$\ldots$}}
\put(253,40){\mbox{$\ldots$}}
\put(239,40){\mbox{$\ldots$}}
\put(225,40){\mbox{$\ldots$}}

\put(222,40){\line(0,-1){10}}
\put(222,30){\line(1,0){10}}
\put(232,30){\line(0,-1){20}}
\put(232,10){\line(1,0){18}}
\put(250,0){\line(0,1){10}}

\put(0,90){\line(1,0){10}}
\put(10,90){\line(0,-1){20}}
\put(10,70){\line(1,0){20}}
\put(30,70){\line(0,-1){10}}
\put(30,60){\line(1,0){10}}
\put(40,60){\line(0,-1){10}}
\put(40,50){\line(1,0){10}}
\put(50,50){\line(0,-1){10}}

{\footnotesize
\put(123,17){\mbox{$ \overline{P}$}}
\put(17,50){\mbox{$R$}}
\put(243,22){\mbox{$\check P$}}
\qbezier(270,3)(280,20)(270,37)
\put(280,18){\mbox{$h_P = l_{P^{\vee}}=p_{_1}$}}
\qbezier(5,-5)(132,-20)(260,-5)
\put(130,-25){\mbox{$N $}}
\qbezier(5,35)(25,25)(45,35)
\put(22,20){\mbox{$l_R$}}
\qbezier(225,43)(245,52)(265,43)
\put(243,52){\mbox{$l_{\!_P}$}}
}

\put(4,40){\mbox{$\ldots$}}
\put(18,40){\mbox{$\ldots$}}
\put(32,40){\mbox{$\ldots$}}

\end{picture}

\noindent
The ordinary $N$-independent representations in this notation are $R=(R,\varnothing)$,
there conjugate are $\overline{R} = (\varnothing,R)$.
The simplest of non-trivial composites is the
adjoint $(1,1) = [2,1^{N-2}]$.

\subsection{Symmetric functions and diagram dependent time variables}

Let us consider a symmetric polynomial of $N$ variables $x_i$ (which are sometimes called Miwa variables). We will mostly consider it as a polynomial of time variables (power sums)
\be
p_k:=\sum_ix_i^k
\ee
In fact, for the proper taking into account the $U(1)$-factor in the Hopf invariants (see a discussion in \cite{tangles,MMhopf}), we should multiply all the $x_i$ by a common factor
\be\label{U(1)}
x_i\to\left(\prod_i x_i\right)^{-\frac{1}{N}}\cdot x_i
\ee
in order to have the product $\prod_ix_i=1$. We will also consider the power sums of inverse Miwa variables, and denote the corresponding quantities as
\be\label{37}
{\bf p}^{*x}_k  =
\left(\prod_{i=1}^N x_i\right)^{-\frac{k}{N}}\cdot \sum_{i=1}^N x_i^k, \ \ \ \ \ \
\bp^{*x}_k  = \left(\prod_{i=1}^N x_i\right)^{\frac{k}{N}}
\cdot\sum_{i=1}^N x_i^{-k}
\ee
From now on, for the sake of brevity, we omit the $U(1)$-factor from the definition of the time variables and, hence, from the Hopf invariants, the factor being easily restored.

Now, one can parameterize the Miwa variables as
\be
x_i:=t^{2i-N-1}\cdot v_i^{-2}
\ee
and $v_i$ are the free parameters.
With this normalization, we have
\be\label{pV}
{\bf p}^{*V}_k = \frac{A^k-A^{-k}}{t^k-t^{-k}}
+ A^{-k} \sum_i t^{(2i-1)k}(v_i^{-2k}-1)\nn\\
\bp^{*V}_k = \frac{A^k-A^{-k}}{t^k-t^{-k}}
+ A^{k} \sum_i t^{(1-2i)k}(v_i^{2k}-1)
\ee
with $A:=t^N$.
Choosing $v_i=q^{\mu_i}$, we will associate further the deformation of the topological locus by a Young diagram $\mu=[\mu_1\geq \mu_2\geq \ldots \geq \mu_{l_\mu}>0]$:
\be
{\bf p}^{*\mu}_k = \frac{A^k-A^{-k}}{t^k-t^{-k}}
+ A^{-k} \sum_i t^{(2i-1)k}(q^{-2k\mu_i}-1)
\ee
and
\be
\bp^{*\mu}_k(A,q,t) = {\bf p}^{*\mu}_k(A^{-1},q^{-1},t^{-1}) =
\frac{A^k-A^{-k}}{t^k-t^{-k}}
+ A^{k} \sum_i t^{-(2i-1)k}(q^{2k\mu_i}-1)
\label{dialocus}
\ee
A lifting of (\ref{dialocus}) to composite representations $\mu\longrightarrow (\mu,\lambda)$
is also straightforward: one associates with $v_i$ the set of $q^{\mu_i}$, $i=1,\ldots,l_{_\mu}$ and $q^{-\lambda_i}$, $i=l_{_\mu}+1,\ldots,l_{_\mu}+l_{_\lambda}$,
\be
{\bf p}^{*(\mu,\lambda)}_k=
{A^k-A^{-k}\over t^k-t^{-k}}+A^{-k}\sum_i t^{(2i-1)k}\Big(q^{-2k\mu_i}-1\Big)
+A^{k}\sum_i t^{(1-2i)k}\Big(q^{2k\lambda_i}-1\Big)
\label{p*comp}
\ee
With the $U(1)$-factor taken into account, this expression would be \cite{compomac}
\be
{\bf p}^{*(\mu,\lambda)}_k=\left(\prod_{i,j} q^{\mu_i-\lambda_j}\right)^{\frac{2k}{N}}\cdot\left(
{A^k-A^{-k}\over t^k-t^{-k}}+A^{-k}\sum_i t^{(2i-1)k}\Big(q^{-2k\mu_i}-1\Big)
+A^{k}\sum_i t^{(1-2i)k}\Big(q^{2k\lambda_i}-1\Big)\right)
\label{pU1}
\ee

\subsection{On Macdonald deformation of the Koike formula\label{compMac}}

The basic special function associated with representation is its character
expressed through the Schur functions,
${\rm Schur}_{(R,\varnothing)}\{{\bf p}\}={\rm Schur}_{R}\{{\bf p}\}$.
For composite representations one needs their generalization, the
composite Schur functions \cite{Koike,Kanno,MMhopf}.
Because of explicit $N$ dependence, they are not easy to define for
arbitrary (generic) values of time-variable ${\bf p}_k$.
Fortunately in most applications we need their reductions to
just $N$-dimensional loci ${\bf p}^{*x}_k$.
At these peculiar locus, the composite Schur functions
can be defined by the uniformization trick of \cite{MMhopf},
and they possess a nice description as a bilinear combination
of {\it skew} Schur functions. For an arbitrary composite representation
 $(R,P)$ made from a pair of Young diagrams $R$ and $P$, there is the Koike formula \cite{Koike}
\be
{\rm Schur}_{(R,P)}\{{\bf p}^{*x}\} =
\sum_{\eta  } (-)^{|\eta|}\cdot {\rm Schur}_{R/\eta}\{{\bf p}^{*x}\}\cdot {\rm Schur}_{P/\eta^\vee}\{\bp^{*x}\}
\label{compoSchur}
\label{Koi}
\ee
Note that $\eta$ in the second factor is transposed. If the $U(1)$-factor, (\ref{U(1)}) is not taken into account, there is an additional factor of $\Big(\prod_i v_i\Big)^{-2l_{{_P}^\vee}}$ on the r.h.s. of (\ref{compoSchur}).

In the case of Macdonald functions, the situation gets more involved.
Uniformization still works, but it provides non-trivial expressions
with additional poles in $A$ in denominators,
e.g. already for the adjoint Macdonald dimension, i.e. value of the corresponding Macdonald polynomial at the
topological locus ${\bf p}^*_k:=\{A^k\}/\{t^k\}$, one gets
$\M_{([1],[1])} = \frac{\{Aq\}\{A\}\{A/t\}}{\{Aq/t\}\{q\}\{t\}}$
instead of naive $\frac{\{Aq\}\{A/t\}}{\{q\}\{t\}}$,
it is this complicated expression which satisfies the uniformization
request
$\left.\M_{([1],[1])}\right|_{A=t^N} = \left.\M_{[2,1^{N-2}]}\right|_{A=t^N}$.

There are three basic modifications of (\ref{compoSchur}) in the Macdonald case:

{\bf (i)} The skew Macdonald polynomials emerge (instead of the skew Schur polynomials) only
in the limit $A\longrightarrow \infty$.

{\bf (ii)} The sum turns into a double sum over arbitrary diagrams $\eta_1$ and $\eta_2$
of equal sizes, but there is no requirement that $\eta_2=\eta_1^\vee$.

{\bf (iii)} The coefficients become rational functions of $q$ and $t$. These coefficients are constructed by the uniformization procedure \cite{uni,CheEl}: one considers the composite representation at various values of $N$, when it is just ordinary representation, and then parameterizes the dependence of coefficients on $N$ via the uniformizing parameter $A:=t^N$. To put it differently,
the composite Macdonald polynomial as a symmetric function of $x_i$ is defined by the stability condition: for $i=1,...,N$
\be\label{Msc}
M_{(R,P)}(x_i)=M_{[\overline{P},R]_N}(x_i)
\ee
where $[\overline{P},R]_N$ is the Young diagram as in the figure of sec.\ref{comp}.
Surprisingly, the dependence on $N$ is given by rational functions of $A$. Thus, generally the coefficients are rational functions of $A$, $q$ and $t$. For instance,
\be\label{cM}
M_{(\varnothing,[1])}(x_i)=M_{[1]}(x_i^{-1})\nn\\
M_{([1],[1])}(x_i)=M_{[1]}(x_i)M_{[1]}(x_i^{-1})-{\{A\}\{q\}\over \{t\}\{Aq/t\}}\nn\\
\ldots
\ee
In other words, (\ref{compoSchur}) is substituted by
\be
M_{(R,P)} \{{\bf p}^{*x}\} = M_{R}\{{\bf p}^{*x}\}\cdot
M_{P}\{\bp^{*x}\}+
\sum_{\stackrel{|\zeta_1|<|R|,|\zeta_2|< |P|}{|R|-|\zeta_1|=|P|-|\zeta_2|}  }
(-1)^{|R|-|\zeta_1|}{\cal B}^{\zeta_1,\zeta_2}_{(R,P)}(A,q,t)
\cdot M_{\zeta_1}\{{\bf p}^{*x}\}\cdot
M_{\zeta_2}\{\bp^{*x}\}
\label{compoMac}
\ee
For instance, in the case of the composite representation $([2,1],[2,1])$, this formula is
{\footnotesize\be
M_{([2,1],[2,1])}\{{\bf p}^{*x}\} =
M_{[2,1]}\{{\bf p}^{*x}\}\cdot M_{[2,1]}\{\bp^{*x}\}
- {\cal B}_{([2,1],[2,1])}^{[2]}\cdot M_{[2]}\{{\bf p}^{*x}\}\cdot M_{[2]}\{\bp^{*x}\}
- {\cal B}_{([2,1],[2,1])}^{[1,1]}\cdot M_{[1,1]}\{{\bf p}^{*x}\}\cdot M_{[1,1]}\{\bp^{*x}\}
- \!\!\!\!\!\!\!\!\!\! \nn \\
\!\!\!\!\!\!\!\!\!\!
-{\cal B}_{([2,1],[2,1])}^{[2],[1,1]}\cdot\Big(M_{[2]}\{{\bf p}^{*x}\}\cdot M_{[1,1]}\{\bp^{*x}\}
+ M_{[1,1]}\{{\bf p}^{*x}\}\cdot M_{[2]}\{\bp^{*x}\}\Big)
+ {\cal B}_{([2,1],[2,1])}^{[1]}\cdot M_{[1]}\{{\bf p}^{*x}\}\cdot M_{[1]}\{\bp^{*x}\}
- {\cal B}_{([2,1],[2,1])}^\varnothing \ \ \ \ \ \ \
\ee}
The coefficients ${\cal B}$ are quite tedious, their manifest expressions can be found in \cite{compomac}.

\section{Refined Hopf link invariant}

\subsection{Hopf link hyperpolynomial}

The refined Hopf link invariant, i.e. Hopf hyperpolynomial is defined by the formula \cite{IK}
\be\label{Hopf}
{\cal P}_{\lambda,\mu}^{\rm Hopf} =\M_\lambda\cdot M_\mu({\bf p}_k^{*\lambda})
\ee
which coincides with the refined version \cite{DMMSS} of the Rosso-Jones formula \cite{RJ}
\be\label{RJ}
{\cal P}_{\lambda,\mu}^{\rm Hopf} =f_\lambda f_\mu
\sum_{\eta\in \lambda\otimes \mu}
{\bf N}_{\lambda\mu}^\eta f_\eta^{-1}
\cdot {\M}_\eta
\ee
Here
\be\label{fr}
f_\mu:=\left(-{q\over t}\right)^{|\mu|}q^{\nu'(\mu)}t^{-\nu(\mu)}=(-1)^{|\mu|}\ q^{\sum\mu_i^2}\ t^{-\sum{\mu^\vee_i}^2}
\ee
is the framing factor \cite{Taki} and
\be\label{nu}
\M_\eta:=M_\eta \{{\bf p}^{*\varnothing}\},\ \ \ \ \ \ \ \nu(\lambda):=2\sum_i (i-1)\lambda_i,\ \ \ \ \ \nu'(\lambda):=\nu(\lambda^\vee)
\ee
The definition (\ref{Hopf}) coincides with \cite{AS,DMMSS,Che}. One can check that ${\cal P}_{\lambda,\mu}^{\rm Hopf}={\cal P}_{\mu,\lambda}^{\rm Hopf}$ as it should be.

Note that in (\ref{Hopf}) we have omitted the $U(1)$-factor $q^{2|\lambda||\mu|\over N}$, \cite{Atiah,MarF,China1,tangles} because of choosing the time variables (\ref{p*comp}) instead of (\ref{pU1}). In (\ref{RJ}), this $U(1)$-factor would be guaranteed by the proper choice of the framing factor: the unit of the unrefined framing factor $q^{C_2(\mu)}$, where $C_2(\mu)$ is the second Casimir operator of $SL_N$ becomes in the refined case
\be
q^{C_2(\mu)}=q^{\nu'(\mu)-\nu(\mu)+|\mu|N-|\mu|^2/N}\longrightarrow \left({q\over t}\right)^{|\mu|}q^{\nu'(\mu)}t^{-\nu(\mu)}
t^{|\mu|N}q^{-|\mu|^2/N}=(-1)^{|\mu|}t^{|\mu|N}q^{-|\mu|^2/N}f_\mu
\ee
The multiplier $(-1)^{|\mu|}t^{|\mu|N}$ drops off (\ref{RJ}) because of the condition $|\lambda|+|\mu|=|\eta|$, and $q^{-|\mu|^2/N}$ becomes $q^{2|\lambda||\mu|\over N}$ in this formula, which is the $U(1)$-factor. Such a choice of framing corresponds to the
the standard, or canonical framing \cite{Atiah,MarF,China1,tangles}\footnote{In practical terms, this framing with the $U(1)$-factor taken into account is characterized by the quasiclassical expansion of the reduced link invariant, $q=e^\hbar$, $t=e^{\beta\hbar}$, $A=e^{N\hbar}$ without the linear term in $\hbar$:
\be
H=1+0\cdot\hbar+O(\hbar^2)
\nn
\ee
}. However, we omit the $U(1)$-factor in order not to overload the formulas.

\subsection{Hopf link hyperpolynomial in the composite representations}

The definition (\ref{Hopf}) can be straightforwardly extended to the composite representations. That is, the Hopf hyperpolynomial with the components colored with two composite representations $(\mu_1,\mu_2)$ and $(\lambda_1,\lambda_2)$, is given by
\be\label{compHopf}
{\cal P}_{(\mu_1,\mu_2),(\lambda_1,\lambda_2)}^{\rm Hopf} =\M_{(\mu_1,\mu_2)}\cdot
M_{(\lambda_1,\lambda_2)}({\bf p}_k^{*(\mu_1,\mu_2)})
\ee
There are two possibilities to evaluate this quantity: one can construct an uniformization of the corresponding composite Macdonald polynomial, see sec.\ref{compMac}, or one can construct an uniformization of the entire Hopf hyperpolynomial. In the latter case, one has to define the Hopf refined invariant in a composite representation $(R,P)$ with the stability condition:
\be
{\cal P}_{(R,P),(S,T)}(A=t^N)={\cal P}_{[\overline{P},R]_N,[\overline{S},T]_N}(A=t^N)
\ee
For instance, in the case of two adjoint representations $([1],[1])$, one constructs the adjoint-adjoint refined invariant by the property ${\cal P}_{([1],[1]),([1],[1])}(A=t^N)={\cal P}_{[2,1^{N-1}],[2,1^{N-1}]}$. This definition coincides with the definition due to I. Cherednik of the DAHA invariant in the composite representation \cite{CheEl} (though they did not consider links and did not consider dimensions). Then, one can obtain (see (\ref{cM}))
\be
{\cal P}_{([1],[1]),([1],[1])}=\M_{([1],[1])}\left( -R_{([1],[1])}+\left[{\Big(qt+(qt)^{-1}\Big)\{Aq/t\}-\{Aq^2/t^2\}\over
\{t\}}\right]^2\right)
\ee
where
\be\label{def}
\M_{([1],[1])}={\{Aq\}\{A\}\{A/t\}\over \{t\}^2\{Aq/t\}},\ \ \ \ \ R_{([1],[1])}:={\{A\}\{q\}\over \{t\}\{Aq/t\}}
\ee

\subsection{$A^{\pm 1}\to 0$ limit of the Hopf hyperpolynomial}

\subsection*{\underline{$A=0$ limit}} Let us discuss what is behaviour of the Hopf hyperpolynomial in the limit of $A\to 0$. Then, the leading behaviour of time variables in (\ref{compHopf}) is
\be\label{48}
{\bf p}_k^{*(\mu,\lambda)}\stackrel{A\to 0}{\approx}{1\over A^k}\left(
-{1\over t^k-t^{-k}}+\sum_i t^{(2i-1)k}\Big(q^{-2k\mu_i}-1\Big)\right)\\
{\bp}_k^{*(\mu,\lambda)}\stackrel{A\to 0}{\approx}{1\over A^k}\left(-{1\over t^k-t^{-k}}
+\sum_i t^{(2i-1)k}\Big(q^{-2k\lambda_i}-1\Big)\right)
\ee
This means that, in (\ref{compoMac}), the contribution of each term in the sum is $A^{-|R|-|P|+|\zeta_1|+|\zeta_2|}$, and the most singular is the first term $M_{R}\{{\bf p}^{*V}\}\cdot M_{P}\{\bp^{*V}\}$. In other words, we obtain
\be
{\cal P}_{(\mu_1,\mu_2),(\lambda_1,\lambda_2)}^{\rm Hopf}\stackrel{A\to 0}{\approx} A^{-|\lambda_1|-|\lambda_2|} \M_{(\mu_1,\mu_2)}W_{\lambda_1,\mu_1}W_{\lambda_2,\mu_2}
\ee
where
\be
W_{\lambda,\mu}:=M_{\lambda}\Big\{-{q^k-q^{-k}\over t^k-t^{-k}}\cdot\overline{p}_k^{(\mu^\vee)}\Big\}
\stackrel{(\ref{id})}{=}(-1)^{|\lambda|}{\overline{h}_{\lambda^\vee}\over h_\lambda}\overline{M}_{\lambda^\vee}\Big\{\overline{p}_k^{(\mu^\vee)}\Big\}
\ee
and we introduced the new time variables that will be of use hereafter,
\be\label{pmu}
p_k^{(\mu)}  = \sum_{i=1}^\infty  q^{2\mu_ik}t^{(1-2i)k} =
\frac{1}{t^{k}-t^{-k}} + \sum_{i=1}^\infty t^{(1-2i)k}(q^{2\mu_ik}-1)=
\frac{1}{t^{k}-t^{-k}}+(q^k-q^{-k})\sum_{i,j\in\mu} t^{(1-2i)k}q^{(2j-1)k}
\ee
so that (\ref{48}) is expressed through them,
\be
-{1\over t^k-t^{-k}}+\sum_i t^{(2i-1)k}\Big(q^{-2k\mu_i}-1\Big)= -{q^k-q^{-k}\over t^k-t^{-k}}\cdot\overline{p}_k^{(\mu^\vee)}=p_k^{(\mu)}\Big|_{q\to q^{-1},t\to  t^{-1}}:=\underline{p_k^{(\mu)}}
\ee
and one can use (\ref{und}).

Similarly,
\be
\M_{(\mu_1,\mu_2)}\stackrel{A\to 0}{\approx} A^{-|\mu_1|-|\mu_2|}W_{\mu_1,\varnothing}W_{\mu_2,\varnothing}=
(-A)^{-|\mu_1|-|\mu_2|}{\overline{h}_{\mu_1^\vee}\over h_{\mu_1}}{\overline{h}_{\mu_2^\vee}\over h_{\mu_2}}
\overline{M}_{\mu_1^\vee}\{\overline{p}^{(\varnothing)}\}\overline{M}_{\mu_2^\vee}\{\overline{p}^{(\varnothing)}\}
\ee
Thus, finally,
\be
{\cal P}_{(\mu_1,\mu_2),(\lambda_1,\lambda_2)}^{\rm Hopf}\stackrel{A\to 0}{\approx}
(-A)^{-|\Sigma|}
{\overline{h}_{\lambda_1^\vee}\overline{h}_{\mu_1^\vee}\over h_{\lambda_1}h_{\mu_1}}
\cdot {\overline{h}_{\lambda_2^\vee}\overline{h}_{\mu_2^\vee}\over h_{\lambda_2}h_{\mu_2}}
\left(\overline{M}_{\lambda_1^\vee}\Big\{\overline{p}_k^{(\mu_1^\vee)}\Big\}\overline{M}_{\mu_1^\vee}\{\overline{p}^{(\varnothing)}\}\right)
\left(\overline{M}_{\lambda_2^\vee}\Big\{\overline{p}_k^{(\mu_2^\vee)}\Big\}\overline{M}_{\mu_2^\vee}\{\overline{p}^{(\varnothing)}\}
\right)
\label{HHA=0}
\ee
Since the expression in the first brackets is the Hopf hyperpolynomial at $A\to 0$ with components colored with $\lambda_1$ and $\mu_1$, it is symmetric w.r.t. permuting $\lambda_1$ and $\mu_1$. Similarly, symmetric is the expression in the second brackets, and the whole formula (\ref{HHA=0}) is clearly symmetric with respect to interchanging $(\mu_1,\mu_2)\leftrightarrow (\lambda_1,\lambda_2)$ as it should be:
\be\label{Hopfas}
\boxed{
{\cal P}_{(\mu_1,\mu_2),(\lambda_1,\lambda_2)}^{\rm Hopf}\ \stackrel{A\to 0}{\approx}\
\left[{\cal P}_{\lambda_1,\mu_1}^{\rm Hopf}\right]_0\cdot
\left[{\cal P}_{\lambda_2,\mu_2}^{\rm Hopf}\right]_0
}
\ee
where we again used formula (\ref{id}) and denoted the Hopf hyperpolynomial at $A\to 0$ as $\left[{\cal P}_{\lambda_2,\mu_2}^{\rm Hopf}\right]_0$.

\subsection*{\underline{$A=\infty$ limit}} Similarly, the $A=\infty$ limit is described by leading behaviour of the time variables
\be
{\bf p}_k^{*(\mu,\lambda)}\stackrel{A\to \infty}{\approx}A^k\left(
{1\over t^k-t^{-k}}+\sum_i t^{(1-2i)k}\Big(q^{2k\lambda_i}-1\Big)\right)\\
{\bp}_k^{*(\mu,\lambda)}\stackrel{A\to \infty}{\approx}A^k\left(
{1\over t^k-t^{-k}}+\sum_i t^{(1-2i)k}\Big(q^{2k\mu_i}-1\Big)\right)
\ee
which means that, in (\ref{compoMac}), the contribution of each term in the sum is $A^{|R|+|P|-|\zeta_1|-|\zeta_2|}$, and the most singular is again the first term $M_{R}\{{\bf p}^{*V}\}\cdot M_{P}\{\bp^{*V}\}$. Performing calculation similar to the previous paragraph, we obtain
\be\label{Hopfas2}
\boxed{
{\cal P}_{(\mu_1,\mu_2),(\lambda_1,\lambda_2)}^{\rm Hopf}\ \stackrel{A\to \infty}{\approx}\ A^{\Sigma}
\left[A^{|\mu_1|+|\lambda_1|}{\cal P}_{\lambda_1,\mu_2}^{\rm Hopf}(A,q^{-1},t^{-1})\right]_0\cdot
\left[A^{|\mu_2|+|\lambda_2|}{\cal P}_{\lambda_2,\mu_1}^{\rm Hopf}(A,q^{-1},t^{-1})\right]_0
}
\ee
where we introduced the notation $\Sigma:=|\mu_1|+|\mu_2|+|\lambda_1|+|\lambda_2|$.
The only difference with (\ref{Hopfas}) is inverting the parameters $q$ and $t$ (and simple matching the degrees of $A$), while the answer still expresses via the behaviour of the ordinary Hopf hyperpolynomials at the vicinity of $A=0$. This is not at all surprising, since they behave at $A\to\infty$ trivially, while the composite Hopf hyperpolynomial does not.

\section{Convolution of refined topological vertices}

\subsection{Refined topological vertex and four point functions}

The original definition of the refined topological vertex, \cite{AK08} was\footnote{Compared with the original convention in \cite{AK08},
the index for the preferred direction has been raised to respect the definition of DIM intertwiner.
Consequently the lower/upper index corresponds to outgoing/incoming arrow in the corresponding diagram.}
\beq\label{rtv}
C_{\lambda}\!\phantom{.}^{\mu\xi} (q,t)  =
f^{-1}_\xi\cdot M_{\mu}\{p^{(\varnothing)}\}
\sum_{\eta} \Big({q\over t}\Big)^{|\eta|-|\xi|}M_{\xi/\eta}\Big\{p_k^{(\mu)}\Big\}\cdot  \overline{M}_{\lambda^\vee/\eta^\vee} \Big\{(-1)^{k+1}\overline{p}_k^{(\mu^\vee)}\Big\}
\eeq
(see also \cite[eqs.(95)]{AKM4OZ}).
We will need also the vertex with raised/lowered indices
\beq
C^{\lambda}_{~~\mu\xi} (q,t) := (-1)^{|\lambda| + |\mu| + |\xi|} C_{\lambda^\vee}^{~~~\mu^\vee\xi^\vee}(t,q)
\eeq

Within the DIM approach, \cite{AFS,AKM4OZ}
the refined topological vertices are just the intertwiners
between a ``horizontal" Fock representation and a tensor product of one ``horizontal" and one ``vertical" Fock representations.

We choose level $(1,0)$ as the canonical choice for the horizontal representation:
\beq
\Psi (v): \mathcal{F}_{v}^{(0,1)} \otimes \mathcal{F}_{u}^{(1,0)} \longrightarrow  \mathcal{F}_{-uv}^{(1,1)}
\nn \\   \\ \nn
\Psi^{*} (v) : \mathcal{F}_{-uv}^{(1,1)}  \longrightarrow  \ \mathcal{F}_{v}^{(0,1)} \otimes \mathcal{F}_{u}^{(1,0)}
\eeq
Then, the topological vertices in these DIM terms are \cite[sec.4.4]{AFS}
(in this section we  mark by red the preferred direction, associated with the vertical representation)\footnote{To match the above diagram with \cite[eqs.(93,94)]{AKM4OZ}, we have to make a reflection of all the edges w.r.t. the vertex (the origin). This does not change the incoming/outgoing direction, but the sign of level (slope) is flipped.}:

\vspace{10mm}

\beq
C_{\lambda}^{~~\mu\nu}(q,t)  \sim  \langle \nu \vert (\Psi^{*})^\mu \vert \lambda \rangle \quad \Longleftrightarrow \quad
\begin{picture}(10,5)(0,-5)
\unitlength 8pt
\thicklines
\put(4,-1){\vector(1,0){3}}
\put(6,-1){\line(1,0){4}}
\put(4,-1){\color{red}\vector(0,1){3}}
\put(4,1){\color{red}\line(0,1){4}}
\put(0,-5){\vector(1,1){2.5}}
\put(2,-3){\line(1,1){2}}
\put(5,0.5){\mbox{$ (\Psi^{*})^\mu$}}
\put(1.5,-5){\mbox{$\lambda$}}
\put(2.5,3.5){\mbox{$\mu$}}
\put(9,-0.5){\mbox{$\nu$}}
\end{picture}
\eeq

\newpage

\vspace{15mm}

\beqa
\begin{picture}(10,5)(0,0)
\put(-200,0){\mbox{$
C^{\lambda}_{~~\mu\nu} (q,t)
\ = \  (-1)^{|\lambda| + |\mu| + |\nu|} C_{\lambda^\vee}^{~~~\mu^\vee\nu^\vee}(t,q)
\ \sim \ \langle \lambda \vert \Psi_\mu \vert \nu \rangle \quad \Longleftrightarrow \quad
$}}
\put(100,0){
\unitlength 8pt
\thicklines
\put(0,0){\vector(1,0){3}}
\put(2,0){\line(1,0){4}}
\put(6,-5){\color{red}\vector(0,1){3}}
\put(6,-3){\color{red}\line(0,1){3}}
\put(6,0){\vector(1,1){2.5}}
\put(8,2){\line(1,1){2}}
\put(7,-0.8){\mbox{$\Psi_\mu$}}
\put(1,1){\mbox{$\nu$}}
\put(6.5,-4){\mbox{$\mu$}}
\put(8,4){\mbox{$\lambda$}}
}
\end{picture}
\eeqa

\vspace{20mm}

\noindent
Note that as a consequence of the chosen position for the preferred direction index, we now have
a complete correspondence; lower/upper index $\Leftrightarrow$ incoming/outgoing arrow.

\subsection{Four point functions by gluing the refined topological vertex}

There are three possible gluings along $(1,1), (1,0)$ and $(0,1)$ directions.
(These slopes can be identified with 5-brane charge.)
The first two are gluing along the unpreferred direction, while the last one is gluing
along the preferred direction, which is qualitatively different from the former gluings.
The gluings along $(1,1)$ and $(1,0)$ directions are geometrically
related by the flop transition (the change of the sign of the K\"ahler parameter),
or the exchange of $\Psi$ and $\Psi^{*}$ from the DIM viewpoint. On the other hand,
there seems no simple relation between the gluings of unpreferred and the preferred
directions. The gluing rules have been checked in  \cite[sec.4.5]{AFS} from the DIM principle. The three different 4-point functions look as follows (the other two 4-point functions which are irrelevant for our purposes here have been considered in \cite{AK08,AFS}):

\bigskip

(1) Gluing along $(1,0)$ direction
\cite[Case 2 (Fig.5)]{AFS}:

\vspace{10mm}
\beqa\label{4p}
{\bf Z}\left[\begin{array}{c|c}\!\! \mu_1& \lambda_2^\vee\\
\lambda_1& \mu_2^\vee\end{array}\right]
 &:=& \sum_{\xi} (-Q)^{|\xi|} C_{\lambda_1}^{~~\mu_1\xi}(q,t)
C_{\lambda_2^\vee}^{~~~ \mu_2^\vee\xi^\vee}(t,q)
\quad \Longleftrightarrow  \quad
\begin{picture}(10,5)(0,-5)
\unitlength 8pt
\thicklines
\put(2,0){\vector(1,0){2,5}}
\put(4,0){\line(1,0){2}}
\put(2,0){\color{red}\vector(0,1){2.5}}
\put(2,2){\color{red}\line(0,1){2}}
\put(-1,-3){\vector(1,1){2}}
\put(0.5,-1.5){\line(1,1){1.5}}
\put(6,-4){\color{red}\vector(0,1){2.5}}
\put(6,-2){\color{red}\line(0,1){2}}
\put(6,0){\vector(1,1){2}}
\put(7.5,1.5){\line(1,1){1.5}}
\put(1.8,-1.5){\mbox{\footnotesize$(\Psi^{*})^{\mu_1}$}}
\put(6.5,-0.5){\mbox{\footnotesize$\Psi_{\mu_2}$}}
\put(4,1){\mbox{\footnotesize$\xi$}}

\put(0.5,3.5){\mbox{\footnotesize$\mu_1$}}
\put(6.5,-3.5){\mbox{\footnotesize$\mu_2$}}

\put(8,3.5){\mbox{\footnotesize$\lambda_2$}}
\put(0,-3.5){\mbox{\footnotesize$\lambda_1$}}

\end{picture}
\CR
&=& \langle \lambda_2 \vert \Psi_{\mu_2}  (\Psi^{*})^{\mu_1} \vert \lambda_1 \rangle
\eeqa
\vspace{10mm}

(2) Gluing along $(1,1)$ direction \cite[Case 3 (Fig.6)]{AFS}:

\vspace{10mm}
\beqa\label{4plambda}
{\bf Z'}\left[\begin{array}{c|c}\mu_1\,\lambda_1&\!\! \\\!\! &\mu_2^\vee\,\lambda_2^\vee\end{array}\right]
&:=& \sum_{\xi} (-Q)^{|\xi|} C_{\xi }^ {~~\mu_1 \lambda_1}(q,t)
C_{\xi^\vee}^{~~~ \mu_2^\vee\lambda_2^\vee}(t,q)
\quad \Longleftrightarrow  \quad
\begin{picture}(10,5)(0,-5)
\unitlength 8pt
\thicklines
\put(5,0){\line(1,1){1.5}}
\put(3.5,-1.5){\vector(1,1){2}}
\put(3.5,-5){\color{red}\vector(0,1){2}}
\put(3.5,-3){\color{red}\line(0,1){1.5}}
\put(0,-1.5){\vector(1,0){2.5}}
\put(2,-1.5){\line(1,0){1.5}}
\put(6.5,1.5){\color{red}\vector(0,1){2}}
\put(6.5,3){\color{red}\line(0,1){2}}
\put(6.5,1.5){\vector(1,0){2}}
\put(8,1.5){\line(1,0){2}}
\put(1.8,-0.7){\mbox{\footnotesize$\Psi_{\mu_2}$}}
\put(6.5,0.2){\mbox{\footnotesize$(\Psi^{*})^{\mu_1}$}}
\put(4.5,1){\mbox{\footnotesize$\xi$}}
\put(7,4.5){\mbox{\footnotesize$\mu_1$}}
\put(1.8,-4.5){\mbox{\footnotesize$\mu_2$}}
\put(9,2){\mbox{\footnotesize$\lambda_1$}}
\put(0,-1){\mbox{\footnotesize$\lambda_2$}}
\end{picture}
\CR
&=& \langle \lambda_1 \vert (\Psi^{*})^{\mu_1}  \Psi_{\mu_2}  \vert \lambda_2 \rangle
\eeqa
\vspace{10mm}

(3) Gluing along $(0,1)$ (the preferred) direction \cite[Case 1 (Fig.4)]{AFS}:

\vspace{10mm}
\beqa\label{4pmu}
{\bf Z''}\left[\begin{array}{c|c}\!\! \lambda_3 & \lambda_4^\vee \\
\lambda_1& \lambda_2^\vee \end{array}\right]&:=& \sum_{\xi} (-Q)^{|\xi|} C_{\lambda_1}^{~~ \xi \lambda_3}(q,t)
C_{\lambda_4^\vee}^{~~ \xi^\vee \lambda_2^\vee}(t,q)
\quad \Longleftrightarrow  \quad
\begin{picture}(10,5)(0,-5)
\unitlength 8pt
\thicklines
\put(5,-2){\color{red}\vector(0,1){2.5}}
\put(5,0){\color{red}\line(0,1){2}}
\put(5,2){\vector(1,1){2}}
\put(6.5,3.5){\line(1,1){1.5}}
\put(2,-5){\vector(1,1){2}}
\put(3,-4){\line(1,1){2}}
\put(5,-2){\vector(1,0){2.5}}
\put(7,-2){\line(1,0){2}}
\put(1,2){\vector(1,0){2.5}}
\put(3,2){\line(1,0){2}}
\put(6,0){\mbox{\tiny{$\displaystyle{\sum_{\xi}}\Psi_\xi \otimes (\Psi^{*})^\xi$}}}
\put(2,-3){\mbox{\footnotesize$\lambda_1$}}
\put(7.5,3.5){\mbox{\footnotesize$\lambda_4$}}
\put(1,3){\mbox{\footnotesize$\lambda_2$}}
\put(8,-3){\mbox{\footnotesize$\lambda_3$}}
\end{picture}
\CR
&=& \langle \lambda_4 \vert \otimes  \langle \lambda_3 \vert
\left[ \displaystyle{\sum_{\xi}}\Psi_\xi \otimes (\Psi^{*})^\xi \right]
\vert  \lambda_2 \rangle \otimes \vert \lambda_1 \rangle
\eeqa
\vspace{10mm}

 The notation is evident: the two lower/upper indices correspond to incoming/outgoing arrows, while the first/second column is associated with the first/second vertex. We also introduced into the definition an additional transposition of the Young diagram entering the second vertex to have some more symmetric expressions.

We discuss below all these three different 4-point functions, in particular, how they can be evaluated (if any). Note that all these functions celebrate the property
\be\label{12}
\widehat{\bf Z}\left[\begin{array}{c|c}\!\! \mu_1& \lambda_2\\
\lambda_1& \mu_2\end{array}\right](A,q,t)=
\widehat{\bf Z}\left[\begin{array}{c|c}\!\! \mu_2& \lambda_1\\
\lambda_2& \mu_1\end{array}\right](Aq/t,t,q)\nn\\
\nn\\
\widehat{\bf Z'}\left[\begin{array}{c|c}\mu_1\,\lambda_1&\!\! \\\!\! &\mu_2\,\lambda_2\end{array}\right](A,q,t)=\widehat{\bf Z'}\left[\begin{array}{c|c}\mu_2\,\lambda_2&\!\! \\\!\! &\mu_1\,\lambda_1\end{array}\right](Aq/t,t,q)\nn\\
\nn\\
\widehat{\bf Z''}\left[\begin{array}{c|c}\!\! \lambda_3 & \lambda_4 \\
\lambda_1& \lambda_2\end{array}\right](A,q,t)={\bf Z''}\left[\begin{array}{c|c}\!\! \lambda_2 & \lambda_1 \\
\lambda_4& \lambda_3\end{array}\right](Aq/t,t,q)
\ee
This property reduces the number of distinct functions in the case when two of the Young diagrams are trivial (see secs.\ref{lit} and 8).

\subsection{Four point function with the internal edge along the unpreferred direction ${\bf Z}$}

We consider in detail the four point function with the internal edge along the unpreferred direction $\xi$. It is
\beq%
{\bf Z}\left[\begin{array}{c|c}\!\! \mu_1& \lambda_2\\
\lambda_1& \mu_2\end{array}\right]
 &:=& \sum_{\xi} (-Q)^{|\xi|} C_{\lambda_1}^{~~\mu_1\xi}(q,t)
C_{\lambda_2}^{~~~ \mu_2\xi^\vee}(t,q)
\eeq
where $Q=A^2q/t$. After using the Cauchy formula (\ref{Cauchy}), it is equal to
\be
{\bf Z}\left[\begin{array}{c|c}\!\! \mu_1& \lambda_2\\
\lambda_1& \mu_2\end{array}\right]
={\bf Z}\left[\begin{array}{c|c}\!\! \varnothing& \varnothing\\
\varnothing& \varnothing\end{array}\right]\cdot
{\cal N}_{(\mu_1,\mu_2)}
\times\sum_{\sigma,\eta_1,\eta_2}\Big(-{A^2q^2\over t^2}\Big)^{|\eta_1|}(-A^2)^{|\eta_2|}\Big(-{A^2q\over t}\Big)^{-|\sigma|}\times
\nn\\
\times\overline{M}_{\lambda_1^\vee/\eta_1}\{(-1)^{k+1}\overline{p}_k^{(\mu_1^\vee)}\}\cdot M_{\lambda_2^\vee/\eta_2}\{(-1)^{k+1}p_k^{(\mu_2^\vee)}\}
\cdot \overline{M}_{\eta_1/\sigma }\{\overline{p}^{(\mu_2)}\}
\cdot M_{\eta_2/\sigma^\vee}\{p^{(\mu_1)}\}
\ee
where ${\cal N}_{(\mu_1,\mu_2)}$ is defined to be
\be\label{60}
{\cal N}_{\mu_1,\mu_2}:=
M_{\mu_1}\{p^{(\varnothing)}\}\cdot
\overline{M}_{\mu_2}\{\overline{p}^{(\varnothing)}\}\cdot{\exp\left(-\sum_k   \frac{Q^k \,p_k^{(\mu_1)}\overline{p}_k^{(\mu_2)}}{k}\right)\over
\exp\left(-\sum_k   \frac{Q^k \,p_k^{(\varnothing)}\overline{p}_k^{(\varnothing)}}{k}\right)}
\ee
Since the $\mu$-independent factor is
\be
\exp\left(-\sum_k   {1\over k}\cdot\frac{Q^k }{(q^k-q^{-k})(t^k-t^{-k})}\right)=\prod_{i,j=1}^\infty (1-Qq^{-2i+1}t^{-2j+1})
\ee
we obtain
\be
\exp\left(-\sum_k   \frac{Q^k \,p_k^{(\mu_1)}\overline{p}_k^{(\mu_2)}}{k}\right)=
{\cal G}_{\mu_1\mu_2}(A,q,t)\cdot
\prod_{i,j=1}^\infty (1-A^2q^{-2i+2}t^{-2j+2})=
\nn\\
=\left(-A{q\over t}\right)^{|\mu_1|}
(-A)^{|\mu_2|}q^{(\nu'(\mu_1)-\nu(\mu_2))/2}t^{(\nu'(\mu_2)-\nu(\mu_1))/2}
h_{\mu_1}\overline{h}_{\mu_2}\times\nn\\
\times\sum_\sigma (-1)^{|\sigma|}
\M_{\mu_1/\sigma}(Aq/t)\cdot\overline{\M}_{\mu_2/\sigma^\vee}\cdot \prod_{i,j=1}^\infty (1-A^2q^{-2i+2}t^{-2j})
\label{Nf}
\ee
where the standard Nekrasov factor
\be
{\cal G}_{\mu\lambda}(A,q,t):=\prod_{(i,j)\in\mu^\vee}\left(1-A^2t^{2\lambda_i-2j}q^{2\mu_j-2i+2}\right)
\prod_{(i,j)\in\lambda}\left(1-A^2t^{-2\mu_i^\vee+2j-2}q^{-2\lambda_j^\vee+2i}\right)
\ee
and
$\M_{\mu/\sigma}$ is obtained by the specialization $p_k={\bf p}^{*\varnothing}_k$ in the skew Macdonald polynomial $M_{\mu/\sigma}\{p_k\}$ and the substitution $A\to Aq/t$ in the argument of this quantity is manifestly indicated.

Now, since
\be\label{Mhook}
M_{\mu}\{p^{(\varnothing)}\}={q^{\nu'(\mu)/2}t^{-\nu(\mu)/2}\over h_\mu(q,t)}
\ee
we finally obtain
\be\label{65}
{\cal N}_{\mu_1,\mu_2}=f_{\mu_1}\overline{f}_{\mu_2}A^{|\mu_1|}\left(A{q\over t}\right)^{|\mu_2|}
\underbrace{\sum_\sigma (-1)^{|\sigma|}
\M_{\mu_1/\sigma}(Aq/t)\cdot \overline{\M}_{\mu_2/\sigma^\vee}}_{D_{(\mu_1,\mu_2)}}\ :=
f_{\mu_1}\overline{f}_{\mu_2}A^{|\mu_1|}\left(A{q\over t}\right)^{|\mu_2|}D_{(\mu_1,\mu_2)}
\ee
$D_{(\mu_1,\mu_2)}$ is not a Macdonald dimension of the composite representation, though it becomes the quantum dimension at $t=q$, as it has to be.
For instance, the Macdonald dimension of the adjoint representation is (see (\ref{def}))
\be
\M_{([1],[1])}={\{Aq\}\{A/t\}\{A\}\over \{t\}^2\{Aq/t\}}
\ee
while
\be\label{Dadj}
D_{([1],[1])}(A,q,t)={\{Aq\}\{A/t\}\over \{t\}\{q\}}
\ee
Note that, similarly to the Macdonald dimensions, $D_{\mu_1,\mu_2}$ is always factorized due to (\ref{Nf}).  Moreover, $D_{(\mu,\varnothing)}=\M_\mu\Big|_{A\to Aq/t}$ and $D_{(\varnothing,\mu)}=\overline{\M}_\mu$.

\bigskip

Now we can use the identity (\ref{Macex}) in order to get
\be
&&\hspace{-1cm}\sum_{\eta_1}\Big(-{A^2q^2\over t^2}\Big)^{|\eta_1|}\cdot
\overline{M}_{\lambda_1^\vee/\eta_1}\{(-1)^{k+1}\overline{p}_k^{(\mu_1^\vee)}\}\cdot \overline{M}_{\eta_1/\sigma }\{\overline{p}_k^{(\mu_2)}\}=\nn\\
&=&\Big({Aq\over t}\Big)^{|\sigma|}\Big(-{Aq\over t}\Big)^{|\lambda_1|}\cdot\sum_{\eta_1}
\overline{M}_{\lambda_1^\vee/\eta_1}\Big\{-\Big({Aq\over t}\Big)^{-k}\overline{p}^{(\mu_1^\vee)}_k\Big\}\cdot
\overline{M}_{\eta_1/\sigma }\Big\{\Big({Aq\over t}\Big)^k\overline{p}^{(\mu_2)}_k\Big\}=\nonumber\\
&=&\Big({Aq\over t}\Big)^{|\sigma|}\Big(-{Aq\over t}\Big)^{|\lambda_1|}\cdot
\overline{M}_{\lambda_1^\vee/\sigma}\Big\{\Big({Aq\over t}\Big)^k\overline{p}^{(\mu_2)}_k-\Big({Aq\over t}\Big)^{-k}\overline{p}^{(\mu_1^\vee)}_k
\Big\}
\ee
Similarly, the second sum is proportional to
\be
M_{\lambda_2^\vee/\sigma^\vee}\{A^{k}p^{(\mu_1)}_k-A^{-k}p^{(\mu_2^\vee)}_k\}
\ee
Thus, restoring all the factors, we finally obtain \cite{AK08}
\be
\boxed{
\begin{array}{c}
\widehat{\bf Z}\left[\begin{array}{c|c}\!\! \mu_1& \lambda_2\\
\lambda_1& \mu_2\end{array}\right]:=
{\bf Z}\left[\begin{array}{c|c}\!\! \mu_1& \lambda_2\\
\lambda_1& \mu_2\end{array}\right]\cdot
{\bf Z}\left[\begin{array}{c|c}\!\! \varnothing& \varnothing\\
\varnothing& \varnothing\end{array}\right]^{-1}
=
(-1)^{|\lambda_1|+|\lambda_2|}A^{\Sigma}\left({q\over t}\right)^{|\mu_2|+|\lambda_1|}f_{\mu_1}\overline{f}_{\mu_2}
D_{(\mu_1,\mu_2)}\times
\cr
\cr
\times
\sum_\sigma (-1)^{|\sigma|}\cdot\overline{M}_{\lambda_1^\vee/\sigma}\Big\{\Big({Aq\over t}\Big)^k\overline{p}^{(\mu_2)}_k-\Big({Aq\over t}\Big)^{-k}\overline{p}^{(\mu_1^\vee)}_k
\Big\}\cdot
M_{\lambda_2^\vee/\sigma^\vee}\{A^{k}p^{(\mu_1)}_k-A^{-k}p^{(\mu_2^\vee)}_k\}
\end{array}
}\nn\\
\label{Z}
\ee
where we introduced the normalized function $\widehat{\bf Z}$ and use this notation for the normalized 4-point functions hereafter.

\subsection{Four point function with the internal edge along the other unpreferred direction ${\bf Z'}$}

Instead of (\ref{4p}), one can convert two refined topological vertices in the other unpreferred direction:
\beq
{\bf Z'}\left[\begin{array}{c|c}\mu_1\,\lambda_1&\!\! \\\!\! &\mu_2\,\lambda_2\end{array}\right]
&:=& \sum_{\xi} (-Q)^{|\xi|} C_{\xi }^ {~~\mu_1 \lambda_1}(q,t)
C_{\xi^\vee}^{~~~ \mu_2\lambda_2}(t,q)
\eeq
One can calculate this expression similarly to the previous paragraph using that
\be
{M_\mu\Big\{p^{(\varnothing)}\Big\}\over \overline{M}_{\mu^\vee}\Big\{\overline{p}^{(\varnothing)}\Big\}}=\left(-{t\over q}\right)^{|\mu|}f_\mu\cdot ||\mu||^2
\ee
In fact, as follows from the manifest form of the vertex (\ref{rtv}), this case is obtained from the previous one
by conjugation of diagrams, changing signs of time variables and permuting $q$ and $t$ in the Macdonald factors in the sum:
\be\label{Zp}
\boxed{
\begin{array}{c}
\widehat{\bf Z'}\left[\begin{array}{c|c}\mu_1\,\lambda_1&\!\! \\\!\! &\mu_2\,\lambda_2\end{array}\right]
=(-1)^{|\mu_1|+|\mu_2|}A^{\Sigma}
\Big({q\over t}\Big)^{|\lambda_2|+|\mu_2|}\overline{f}_{\lambda_2}^{-1}f^{-1}_{\lambda_1}\cdot ||\mu_1||^2 ||\mu_2||^2\cdot D_{(\mu_2^\vee,\mu_1^\vee)}\times\cr\cr
\times\sum_\sigma (-1)^{|\sigma|}\cdot M_{\lambda_1/\sigma}\Big\{-\Big({Aq\over t}\Big)^k p^{(\mu_2^\vee)}_k+\Big({Aq\over t}\Big)^{-k} p^{(\mu_1)}_k
\Big\}\cdot
\overline{M}_{\lambda_2/\sigma^\vee}\{-A^{k}\overline{p}^{(\mu_1^\vee)}_k+A^{-k}\overline{p}^{(\mu_2)}_k\}
\end{array}}
\ee
This means that the relation of the two 4-point functions is manifestly described by the formula
\be\label{flop}
\widehat{\bf Z'}\left[\begin{array}{c|c}\mu_1\,\lambda_1&\!\! \\\!\! &\mu_2\,\lambda_2\end{array}\right](A,q,t)=\mathfrak{N}\cdot
\widehat{\bf Z}\left[\begin{array}{c|c}\!\! \mu_2& \lambda_2^\vee\\
\lambda_1^\vee& \mu_1\end{array}\right](A^{-1},t,q)
\ee
where the framing coefficient is
\be
\mathfrak{N}=\left(A^2{q\over t}\right)^{\Sigma}
{(-1)^{|\lambda_1|+|\lambda_2|}
\over f_{\mu_1}\overline{f}_{\mu_2}f_{\lambda_1}\overline{f}_{\lambda_2}}
\ee
The equality (\ref{flop}) is related to the flop operation discussed in \cite{AK08} with an additional transformation, (\ref{12}).

\subsection{Four point function with the internal edge along the preferred direction ${\bf Z''}$}

Instead of (\ref{4p}), one can also convert two refined topological vertices in the preferred direction:
\beq
{\bf Z''}\left[\begin{array}{c|c}\!\! \lambda_3 & \lambda_4 \\
\lambda_1& \lambda_2\end{array}\right]&:=& \sum_{\xi} (-Q)^{|\xi|} C_{\lambda_1}^{~~ \xi \lambda_3}(q,t)
C_{\lambda_4}^{~~ \xi^\vee \lambda_2}(t,q)
\eeq
However, this case is hard to deal with, because one can no longer use the Cauchy formula (\ref{Cauchy}). What one can do is to calculate this sum term by term in powers of $Q$. Since the normalized 4-point function is a finite degree polynomial of $A$ \cite{AK09}, at any concrete representation it can be manifestly calculated.

One can, however, calculate this four point function in the special case of two empty Young diagrams:
\be
{\bf Z''}\left[\begin{array}{c|c}\!\! \lambda_1 & \varnothing \\
\varnothing& \lambda_2\end{array}\right]=f^{-1}_{\lambda_1}\overline{f}^{-1}_{\lambda_2}
  \Big({q\over t}\Big)^{|\lambda_2|-|\lambda_1|}\sum_\xi (-Q)^{|\xi|}
M_{\xi}\{p^{(\varnothing)}\}\overline{M}_{\xi^\vee}\{\overline{p}^{(\varnothing)}\}
M_{\lambda_1}\Big\{p_k^{(\xi)}\Big\}\overline{M}_{\lambda_2}\Big\{\overline{p}_k^{(\xi^\vee)}\Big\}=\nn\\
 \ \stackrel{(\ref{77})}{=} \ f^{-1}_{\lambda_1}\overline{f}^{-1}_{\lambda_2}M_{\lambda_1}\{p^{(\varnothing)}\}
\overline{M}_{\lambda_2}\{\overline{p}^{(\varnothing)}\}  \Big({q\over t}\Big)^{|\lambda_2|-|\lambda_1|}
\sum_\xi (-Q)^{|\xi|}M_{\xi}\Big\{p_k^{(\lambda_1)}\Big\}\overline{M}_{\xi^\vee}\Big\{\overline{p}_k^{(\lambda_2)}\Big\}=\nn\\
=f^{-1}_{\lambda_1}\overline{f}^{-1}_{\lambda_2}M_{\lambda_1}\{p^{(\varnothing)}\}
\overline{M}_{\lambda_2}\{\overline{p}^{(\varnothing)}\}  \Big({q\over t}\Big)^{|\lambda_2|-|\lambda_1|}
\exp\left(-\sum_k   \frac{Q^k \,p_k^{(\lambda_1)}\overline{p}_k^{(\lambda_2)}}{k}\right)
\ee
where we used the formula \cite[sec.6,eq.(6.6)]{Mac} (see also \cite{IK})
\be\label{77}
{M_\mu(p^{(\lambda)})\over M_\mu(p^{(\varnothing)})}={M_\lambda(p^{(\mu)})\over M_\lambda(p^{(\varnothing)})}
\ee
Thus, we arrive at (see (\ref{60}))
\be
\widehat{\bf Z''}\left[\begin{array}{c|c}\!\! \lambda_1 & \varnothing \\
\varnothing& \lambda_2\end{array}\right]=
\Big({q\over t}\Big)^{|\lambda_2|-|\lambda_1|} f^{-1}_{\lambda_1}\overline{f}^{-1}_{\lambda_2}
{\cal N}_{\lambda_1,\lambda_2}
\ee
and, using (\ref{65}), we finally obtain
\be\label{79}
\widehat{\bf Z''}\left[\begin{array}{c|c}\!\! \lambda_1 & \varnothing \\
\varnothing& \lambda_2\end{array}\right]=\Big({q\over t}\Big)^{2|\lambda_2|-|\lambda_1|} A^{|\lambda_1|+|\lambda_2|}D_{(\lambda_1,\lambda_2)}
\ee
A similar calculation for the particular case of  ${\bf Z''}\left[\begin{array}{c|c}\!\! \varnothing & \lambda_2 \\
\lambda_1& \varnothing\end{array}\right]$,
\be\label{80}
{\bf Z''}\left[\begin{array}{c|c}\!\! \varnothing & \lambda_2 \\
\lambda_1& \varnothing\end{array}\right]=(-1)^{|\lambda_1|+|\lambda_2|}\sum_\xi (-Q)^{|\xi|}
M_{\xi}\{p^{(\varnothing)}\}\overline{M}_{\xi^\vee}\{\overline{p}^{(\varnothing)}\}
\overline{M}_{\lambda_1^\vee}\Big\{-\overline{p}_k^{(\xi^\vee)}\Big\}M_{\lambda_2}\Big\{-p_k^{(\xi)}\Big\}
\ee
is not immediate because of the minus signs in the arguments of the Macdonald polynomials so that formula (\ref{77}) can not be used. We will return to this issue in sec.\ref{lit}.

\section{Hopf hyperpolynomial versus refined topological 4-point functions\label{s6}}

As we discussed in sec.\ref{firlev}, the Hopf hyperpolynomial in composite representations is not equal to any of the refined topological 4-point functions. However, it is instructive to compare their $A^{\pm 1}\to 0$ behaviours in order to see they are quite similar, the only difference being that the topological 4-point functions are {\it non-chiral} products of two factors with permuted $q$ and $t$, while the Hopf hyperpolynomial is a product of the same factors but of the same {\it chirality}.

\subsection*{\underline{$A\to 0$ limit}}

In this limit, one has to leave only the first term in each sum over $\xi$ in (\ref{4p}), (\ref{4plambda}), and (\ref{4pmu}).
This means that
\be
{\bf Z}\left[\begin{array}{c|c}\!\! \mu_1& \lambda_2\\
\lambda_1& \mu_2\end{array}\right]\stackrel{A=0}{\approx} M_{\mu_1}\{p^{(\varnothing)}\}
\overline{M}_{\lambda_1^\vee}\{-\overline{p}^{(\mu_1^\vee)}\}\overline{M}_{\mu_2}\{\overline{p}^{(\varnothing)}\}
M_{\lambda_2^\vee}\{-p^{(\mu_2^\vee)}\}
\label{ZA0}
\ee
and similarly for ${\bf Z'}$. However, a finite sum remains in ${\bf Z''}$:
\be
{\bf Z''}\left[\begin{array}{c|c}\!\! \lambda_3 & \lambda_4 \\
\lambda_1& \lambda_2\end{array}\right]
\stackrel{A=0}{\approx} f_{\lambda_3}^{-1}\overline{f}_{\lambda_2}^{-1}
\sum_{\eta_1,\eta_2} \Big(-{q\over t}\Big)^{|\eta_1|-|\lambda_3|+|\lambda_2|-|\eta_2|}
\cdot \nn \\ \cdot
M_{\lambda_3/\eta_1}\Big\{p_k^{(\varnothing)}\Big\}\cdot  \overline{M}_{\lambda_1^\vee/\eta_1^\vee} \Big\{-\overline{p}_k^{(\varnothing)}\Big\}\cdot
\overline{M}_{\lambda_2/\eta_2}\Big\{\overline{p}_k^{(\varnothing)}\Big\}\cdot  M_{\lambda_4^\vee/\eta_2^\vee} \Big\{-p_k^{(\varnothing)}\Big\}
\\ \nn
\label{ZppA0}
\ee
For small representations, these expressions can be compared with the Hopf hyperpolynomial: for instance, for the two adjoint representations they are equal to
\be
{\bf Z}\left[\begin{array}{c|c}\!\! [1]& [1]\cr
[1]& [1]\end{array}\right]_{A=0}={\bf Z''}\left[\begin{array}{c|c}\!\! [1] & [1] \cr
[1]& [1]\end{array}\right]_{A=0}={q^2-1+t^{-2}\over\{t\}^2}\cdot{t^2-1+q^{-2}\over\{q\}^2}
\ee
while the Hopf hyperpolynomial in this case has the $A\to 0$ asymptotics
\be
\left[{\cal P}_{([1],[1]),([1],[1])}^{\rm Hopf}\right]_0=A^{-4}\Biggr({t^2\over q^2}\cdot{q^2-1+t^{-2}\over\{t\}^2}\Biggr)^2
\ee
It is clear that, in the topological case, the answer is a product of two factors, which exchange upon permuting $q$ and $t$, while the Hopf hyperpolynomial is a square of one of these terms. One may say the Hopf hyperpolynomial has a kind of ``chiral structure'', while the topological amplitude corresponds to ``non-chiral'' combination.

In higher representations, such a simple relation with the Hopf hyperpolynomials disappears. It is, however, present in ${\bf Z'}$, which can always be expressed in the $A=0$ limit through the Hopf hyperpolynomials\footnote{Note that $\left[{\cal P}_{\lambda_2^\vee, \mu_2^\vee}^{\rm Hopf}(q,t)\right]_0=
(-1)^{|\mu|+|\lambda|}\cdot||\lambda||^{-2}\cdot||\mu||^{-2}\cdot\left[{\cal P}_{\lambda_2,\mu_2}^{\rm Hopf}(t^{-1},q^{-1})\right]_0$.}:
\be
\boxed{
\begin{array}{c}
{\bf Z'}\left[\begin{array}{c|c}\mu_1\,\lambda_1&\!\! \\\!\! &\mu_2\,\lambda_2\end{array}\right]\stackrel{A=0}{\approx}\left({q\over t}\right)^{|\lambda_2|-|\lambda_1|}
f_{\lambda_1}^{-1}\overline{f}_{\lambda_2}^{-1}\cdot
M_{\mu_1}\{p^{(\varnothing)}\}M_{\lambda_1}\{p^{(\mu_1)}\}\overline{M}_{\mu_2}\{\overline{p}^{(\varnothing)}\}
\overline{M}_{\lambda_2}\{\overline{p}^{(\mu_2)}\}=
\cr
\cr
=A^{\Sigma}\cdot
\left({q\over t}\right)^{|\lambda_2|-|\lambda_1|}
f_{\lambda_1}^{-1}\overline{f}_{\lambda_2}^{-1}\cdot
\left[{\cal P}_{\lambda_1,\mu_1}^{\rm Hopf}(A,q^{-1},t^{-1})\right]_0\cdot
\left[{\cal P}_{\lambda_2,\mu_2}^{\rm Hopf}(A,t^{-1},q^{-1})\right]_0
\end{array}
}
\label{ZpA0}
\ee
We still observe the same phenomenon: this expression is a product of two terms of different ``chirality'', while the corresponding Hopf hyperpolynomial asymptotics (\ref{Hopfas}) is a product of  two terms of the same ``chirality''. In other words, the answers in this case would coincide if one permutes $q$ and $t$ in one of the factors.

\subsection*{\underline{$A\to \infty$ limit}}

This limit of the topological 4-point functions is not as simple as their $A\to 0$ limit. In the case of ${\bf Z}$ and ${\bf Z'}$, one use the manifest expressions (\ref{Z}) and (\ref{Zp}) accordingly. For instance, one can immediately obtains from (\ref{Z})
\be
\boxed{
\begin{array}{c}
\widehat{\bf Z}\left[\begin{array}{c|c}\!\! \mu_1& \lambda_2\\
\lambda_1& \mu_2\end{array}\right]\stackrel{A=\infty}{\approx}
\left(A^2{q\over t}\right)^{\Sigma}
\left(-{q\over t}\right)^{|\lambda_1|-|\lambda_2|}f_{\mu_1}\overline{f}_{\mu_2}
\cdot M_{\mu_1}\{p^{(\varnothing)}\}\cdot\overline{M}_{\mu_2}\{\overline{p}^{(\varnothing)}\}
\cdot\overline{M}_{\lambda_1^\vee}\Big\{\overline{p}^{(\mu_2)}
\Big\}\cdot
M_{\lambda_2^\vee}\Big\{p^{(\mu_1)}\Big\}=
\cr
\cr
=(-A^2)^{\Sigma}\left({q\over t}\right)^{2|\lambda_1|}
\left(-{q\over t}\right)^{|\mu_1|+|\mu_2|}f_{\mu_1}\overline{f}_{\mu_2}
\left[A^{|\mu_1|+|\lambda_1|}{\cal P}_{\lambda_1^\vee,\mu_2}^{\rm Hopf}(A,t^{-1},q^{-1})\right]_0\cdot
\left[A^{|\mu_2|+|\lambda_2|}{\cal P}_{\lambda_2^\vee,\mu_1}^{\rm Hopf}(A,q^{-1},t^{-1})\right]_0
\end{array}}
\label{Zi}
\ee
which has to be compared with (\ref{Hopfas2}).

At the same time, from (\ref{Zp}) it follows that
\be
\widehat{\bf Z'}\left[\begin{array}{c|c}\mu_1\,\lambda_1&\!\! \\\!\! &\mu_2\,\lambda_2\end{array}\right]\stackrel{A=\infty}{\approx}
\left(-A^2{q\over t}\right)^{\Sigma}{1\over f_{\mu_1}\overline{f}_{\mu_2}f_{\lambda_1}\overline{f}_{\lambda_2}}
\cdot M_{\mu_1}\{p^{(\varnothing)}\}\cdot\overline{M}_{\mu_2}\{\overline{p}^{(\varnothing)}\}
\cdot M_{\lambda_1}\Big\{-p^{(\mu_2^\vee)}\Big\}\cdot\overline{M}_{\lambda_2}\Big\{-\overline{p}^{(\mu_1^\vee)}
\Big\}
\label{Zpi}
\ee
and the comparison with the Hopf hyperpolynomial asymtptotics is not immediate.

Thus, we observe that only ${\bf Z'}$ at $A\to 0$ and ${\bf Z}$ at $A\to\infty$ can be directly compared with the asymptotics of the Hopf hyperpolynomial, while comparison of the other 4-point functions with the Hopf asymptotics is not that immediate because of minus signs in the arguments of Macdonald polynomials, and requires an additional changing basis. We discuss this issue in the next section.

\section{Topological vertex and Hopf invariant: changing basis
\label{lit}}

\subsection{Hopf hyperpolynomial and the 4-point function with 2 non-trivial Young diagrams}

\paragraph{The sum over unpreferred direction.} The relation of the Hopf hyperpolynomial with the topological 4-point function with only two non-trivial Young diagrams and the internal edge along the non-preferred direction is known since the paper \cite{AK08} and can be easily obtained from formulas (\ref{Z}), (\ref{Zp}). Indeed:
\begin{itemize}
\item One can put $\mu_2=\lambda_1=\varnothing$  (or $\mu_1=\lambda_2=\varnothing$) in $\widehat{\bf Z}\left[\begin{array}{c|c}\!\! \mu_1& \lambda_2\\
\lambda_1& \mu_2\end{array}\right]$ in order to get formula (\ref{7}), or to put $\mu_1=\lambda_1=\varnothing$  (or $\mu_2=\lambda_2=\varnothing$) in $\widehat{\bf Z'}\left[\begin{array}{c|c}\mu_1\,\lambda_1&\!\! \\\!\! &\mu_2\,\lambda_2\end{array}\right]$ in order to get (\ref{8}).
\item One can also put $\lambda_1=\lambda_2=\varnothing$ in $\widehat{\bf Z}\left[\begin{array}{c|c}\!\! \mu_1& \lambda_2\\
\lambda_1& \mu_2\end{array}\right]$ and $\widehat{\bf Z'}\left[\begin{array}{c|c}\mu_1\,\lambda_1&\!\! \\\!\! &\mu_2\,\lambda_2\end{array}\right]$ in order to get
\be\label{ZD}
\widehat{\bf Z}\left[\begin{array}{c|c}\!\! \mu_1& \varnothing\\
\varnothing& \mu_2\end{array}\right]=
(-A)^{|\mu_1|+|\mu_2|}\left({q\over t}\right)^{|\mu_1|}f_{\mu_1}\overline{f}_{\mu_2}
D_{(\mu_1,\mu_2)}\\
\widehat{\bf Z'}\left[\begin{array}{c|c}\mu_1\,\varnothing&\!\! \\\!\! &\mu_2\,\varnothing\end{array}\right]=(-A)^{|\mu_1|+|\mu_2|}\left({q\over t}\right)^{|\mu_2|}||\mu_1||^2
||\mu_2||^2D_{(\mu_2^\vee,\mu_1^\vee)}
\ee
As we already mentioned earlier, these quantities though being products of simple factors are still not associated with hyperpolynomials of the unknot, i.e. with Macdonald dimensions of the composite representations.
\item There is also the third possibility: one can put $\mu_1=\lambda_1=\varnothing$  (or $\mu_2=\lambda_2=\varnothing$) in $\widehat{\bf Z}\left[\begin{array}{c|c}\!\! \mu_1& \lambda_2\\
\lambda_1& \mu_2\end{array}\right]$, or to put $\mu_2=\lambda_1=\varnothing$  (or $\mu_1=\lambda_2=\varnothing$)  in $\widehat{\bf Z'}\left[\begin{array}{c|c}\mu_1\,\lambda_1&\!\! \\\!\! &\mu_2\,\lambda_2\end{array}\right]$. In this case, one obtains an additional minus sign in the arguments of the Macdonald polynomials in (\ref{Z}) and (\ref{Zp}), which does not allow one to associate it with the Hopf hyperpolynomial:
\be
\widehat{\bf Z}\left[\begin{array}{c|c}\!\! \varnothing& \lambda\\
\varnothing& \mu\end{array}\right]=
(-A)^{|\mu|+|\lambda|}\overline{f}_{\mu}\cdot
\overline{\M}_{\mu}
\cdot M_{\lambda^\vee}\Big\{-{\bf p}^{*\mu^\vee}_k(q^{-1},t^{-1})
\Big\}
\ee
where we used that
\be
A^k\overline{p}_k^{(\varnothing)}-A^{-k}p^{(\mu^\vee)}_k=
-{\bf p}^{*\mu^\vee}_k(q^{-1},t^{-1})
\ee

Let us note, however, that changing the basis would correct this problem. Indeed, consider the open string partition function with two holonomies $U_1$, $U_2$, the corresponding time variables being $\xi^{(1)}_k:=\tr U_1^k$, $\xi^{(2)}_k:=\tr U_2^k$:
\be
Z_{str}(0,U_1,0,U_2):=\sum_{\mu,\lambda}
\widehat{\bf Z}\left[\begin{array}{c|c}\!\! \varnothing& \lambda\\
\varnothing& \mu\end{array}\right]\cdot M_{\mu}(\xi^{(1)})\cdot
\overline{M}_{\lambda}(\xi^{(2)})=\nn\\
=\sum_{\mu}(-A)^{|\mu|}\overline{f}_{\mu}\cdot
\overline{\M}_{\mu}\cdot
\exp\left(\sum_k{A^k\over k}\cdot {\bf p}^{*\mu^\vee}_k(q^{-1},t^{-1})\cdot
\xi^{(2)}_k\right)\cdot M_{\mu}(\xi^{(1)})
\ee
This partition can be transformed to
\be
Z_{str}(0,U_1,0,U_2)=\sum_{\mu,\lambda}(-A)^{|\mu|+|\lambda|}\overline{f}_{\mu}
\cdot\overline{\M}_{\mu}\cdot
M_{\lambda^\vee}\Big\{{\bf p}^{*\mu^\vee}_k(q^{-1},t^{-1})\Big\}
\cdot M_{\mu}(\xi^{(1)})\cdot \overline{M}_{\lambda}\{-\xi^{(2)}\}=\nn\\
=\sum_{\mu,\lambda}(-1)^{|\mu|}A^{|\mu|+|\lambda|}\overline{f}_{\mu}
{\overline{h}_\lambda\over h_{\lambda^\vee}}\cdot{\cal P}^{\rm Hopf}_{\mu,\lambda}(A,t,q)\cdot M_{\mu}(\xi^{(1)})
\cdot \overline{M}_{\lambda}\{-\xi^{(2)}\}
\ee
\item In the last possible case of $\mu_1=\mu_2=\varnothing$, one can again get the $D_{(\mu_1,\mu_2)}$ like (\ref{ZD}) after changing the basis:
\be
Z_{str}(0,0,U_1,U_2)=\sum_{\lambda_1,\lambda_2}\widehat{\bf Z}\left[\begin{array}{c|c}\!\! \varnothing& \lambda_2\\
\lambda_1& \varnothing\end{array}\right]
M_{\lambda_1}\Big\{\xi^{(3)}\Big\}
\overline{M}_{\lambda_2}\Big\{\xi^{(4)}\Big\}=\nn\\
=\sum_{\lambda_1,\lambda_2}A^{|\lambda_1|+|\lambda_2|}\left({q\over t}\right)^{|\lambda_1|}
||\lambda_1||^{-2}\cdot\overline{||\lambda_2||}^{-2}\cdot D_{(\lambda_1,\lambda_2)}\cdot
M_{\lambda_1}\Big\{-\xi^{(3)}\Big\}
\overline{M}_{\lambda_2}\Big\{-\xi^{(4)}\Big\}
\ee
\end{itemize}

\paragraph{Sum over the preferred direction.}
A similar trick with changing the basis would help in the case of the internal edge along the preferred direction, (\ref{4pmu}). First of all note that there are 4 essentially different cases for $\widehat{\bf Z''}\left[\begin{array}{c|c}\!\! \lambda_3 & \lambda_4 \\
\lambda_1& \lambda_2\end{array}\right]$ with two non-empty Young diagrams: $\lambda_1=\lambda_4=\varnothing$, $\lambda_3=\lambda_4=\varnothing$, $\lambda_2=\lambda_3=\varnothing$ and $\lambda_2=\lambda_4=\varnothing$. Let us consider all of them:
\begin{itemize}
\item The first case was considered in (\ref{79}) and is equal to $D_{(\lambda_2,\lambda_3)}$.
\item The second case reduces to the formulas obtained in \cite{IK}. Indeed, they obtained
\be
\underline{\widehat{\bf Z}}''\left[\begin{array}{c|c}\!\! \mu & \lambda \\
\varnothing& \varnothing\end{array}\right](A,q,t)= (-A)^{|\lambda|}\left(-A{q\over t}\right)^{|\mu|}f_\mu \overline{f}_\mu \cdot
{\cal P}^{\rm Hopf}_{\mu,\lambda}(A^{-1},t^{-1},q^{-1})
\ee
However, the definition of the topological vertex in \cite{IK} differs from (\ref{rtv}) and is not consistent with the DIM algebra, it is defined as
\be
\underline{C}_{\lambda}\!\phantom{.}^{\mu\xi} (q,t)  =
f^{-1}_\xi\cdot M_{\mu}\{p^{(\varnothing)}\}
\sum_{\eta} \Big({q\over t}\Big)^{|\eta|-|\xi|}M_{\xi/\eta}\Big\{p_k^{(\mu)}\Big\}\cdot  \overline{M}_{\lambda^\vee/\eta^\vee} \Big\{(-1)^{k}\overline{p}_k^{(\mu^\vee)}\Big\}
\ee
There is an additional minus sign in the argument of the last Macdonald polynomial in this formula as compared with (\ref{rtv}), and it is essential for the representations at higher levels, it does not influence only the fundamental representation, hence (\ref{9}). Thus, the additional minus sign would again prevent associating the topological 4-point function with the Hopf hyperpolynomial. Hence, we can again repeat the trick with changing the basis:
\be
Z''_{str}(0,U_1,U_2,0)=\sum_{\mu,\lambda}\widehat{\bf Z''}\left[\begin{array}{c|c}\!\! \mu & \lambda \\
\varnothing& \varnothing\end{array}\right](A,q,t)\cdot
\overline{M}_{\mu}\Big\{\xi^{(1)}\Big\}\overline{M}_{\lambda^\vee}\Big\{\xi^{(2)}\Big\}\sim\nn\\
\sim\sum_{\mu,\lambda}(-Q)^{|\sigma|}f^{-1}_\lambda\left({q\over t}\right)^{|\lambda|}\cdot
M_{\sigma}\Big\{p^{(\varnothing)}\Big\}\overline{M}_{\sigma^\vee}\Big\{\overline{p}^{(\varnothing)}\Big\}
M_{\mu^\vee}\Big\{(-1)^{k+1}p_k^{(\sigma)}\Big\}M_{\lambda}\Big\{p_k^{(\sigma)}\Big\}\cdot
\overline{M}_{\mu}\Big\{\xi^{(1)}\Big\}\overline{M}_{\lambda^\vee}\Big\{\xi^{(2)}\Big\}=\nn\\
=\sum_{\mu,\lambda}(-Q)^{|\sigma|}f^{-1}_\lambda\left({q\over t}\right)^{|\lambda|}\cdot
M_{\sigma}\Big\{p^{(\varnothing)}\Big\}\overline{M}_{\sigma^\vee}\Big\{\overline{p}^{(\varnothing)}\Big\}
M_\lambda\Big\{p^{(\sigma)}\Big\}M_{\mu^\vee}\Big\{(-1)^{k}p_k^{(\sigma)}\Big\}\cdot
\overline{M}_{\mu}\Big\{-\xi^{(1)}\Big\}\overline{M}_{\lambda^\vee}\Big\{-\xi^{(2)}\Big\}\sim\nn\\
\sim\sum_{\mu,\lambda}\ f^{-1}_{\lambda}\left(-{q\over t}\right)^{2|\mu|+|\lambda|}A^{|\mu|+|\lambda|}\cdot {\cal P}^{\rm Hopf}_{\mu,\lambda^\vee}(A^{-1},t^{-1},q^{-1})\cdot
\overline{M}_{\mu}\Big\{-\xi^{(1)}\Big\}\overline{M}_{\lambda^\vee}\Big\{-\xi^{(2)}\Big\}
\ee
where we omitted the normalization factor ${{\bf Z''}\left[\begin{array}{c|c}\!\! \varnothing & \varnothing \\
\varnothing& \varnothing\end{array}\right]}^{-1}$ from the intermediate expressions.
\item The third case is described by (\ref{80}). One can similarly remove the minus signs in (\ref{80}) by changing the basis, which would immediately allow one to evaluate it:
\be
Z''_{str}(U_1,U_2,0,0)=\sum_{\lambda_1,\lambda_2}\widehat{\bf Z''}\left[\begin{array}{c|c}\!\! \varnothing & \lambda_2 \\
\lambda_1& \varnothing\end{array}\right]
\cdot M_{\lambda_1}\Big\{\xi^{(1)}\Big\}\overline{M}_{\lambda_2}\Big\{\xi^{(2)}\Big\}=\nn\\
=\sum_{\lambda_1,\lambda_2}A^{|\lambda_1|+|\lambda_2|}\left({q\over t}\right)^{2|\lambda_2|-|\lambda_1|}
D_{(\lambda_1,\lambda_2)}
\cdot M_{\lambda_1}\Big\{-\xi^{(1)}\Big\}\overline{M}_{\lambda_2}\Big\{-\xi^{(2)}\Big\}
\ee
\item At last, in the fourth possible case of $\widehat{\bf Z''}\left[\begin{array}{c|c}\!\! \lambda_3 & \varnothing \\
\lambda_1& \varnothing\end{array}\right]$, it is difficult to present summation whatever the signs in the arguments of the Macdonald polynomials are.
\end{itemize}

\subsection{Topological 4-point function and the Hopf hyperpolynomial in the $A^{\pm 1}\to 0$ limit}

As we observed in sec.\ref{s6}, in the limit of $A\to 0$, only one of the 4-point functions, $\widehat{\bf Z}'$ is associated with the product of the Hopf hyperpolynomials in this limit, (\ref{ZpA0}). The reason is again an additional minus sign in the argument of the Macdonald polynomials in (\ref{ZA0}) and (\ref{ZppA0}). Hence, we can again change the basis in order to establish a link to the Hopf hyperpolynomial in this limit. For instance, one can introduce the generating function
\be
Z_{str}(U_1,U_2,U_3,U_4)=\sum_{\mu_1,\mu_2,\lambda_1,\lambda_2}
\widehat{\bf Z}\left[\begin{array}{c|c}\!\! \mu_1& \lambda_2\\
\lambda_1& \mu_2\end{array}\right]\overline{M}_{\mu_1^\vee}\Big\{\xi^{(1)}\Big\}
M_{\mu_2^\vee}\Big\{\xi^{(2)}\Big\}M_{\lambda_1}\Big\{\xi^{(3)}\Big\}
\overline{M}_{\lambda_2}\Big\{\xi^{(4)}\Big\}
\ee
and re-expand it into the basis with changed signs of the time variables $\xi_k^{(3,4)}$:
\be
Z_{str}(U_1,U_2,U_3,U_4)=\sum_{\mu_1,\mu_2,\lambda_1,\lambda_2}
\widetilde{\bf Z}\left[\begin{array}{c|c}\!\! \mu_1& \lambda_2\\
\lambda_1& \mu_2\end{array}\right]\overline{M}_{\mu_1^\vee}\Big\{\xi^{(1)}\Big\}
M_{\mu_2^\vee}\Big\{\xi^{(2)}\Big\}M_{\lambda_1}\Big\{-\xi^{(3)}\Big\}
\overline{M}_{\lambda_2}\Big\{-\xi^{(4)}\Big\}
\ee
This gives:
\be
\widetilde{\bf Z}\left[\begin{array}{c|c}\!\! \mu_1& \lambda_2\\
\lambda_1& \mu_2\end{array}\right]\stackrel{A=0}{\approx} M_{\mu_1}\{p^{(\varnothing)}\}
\overline{M}_{\lambda_1^\vee}\{\overline{p}^{(\mu_1^\vee)}\}\overline{M}_{\mu_2}\{\overline{p}^{(\varnothing)}\}
M_{\lambda_2^\vee}\{p^{(\mu_2^\vee)}\}=\nn\\
\nn\\
=(-A)^{\Sigma}\left(-{q\over t}\right)^{|\mu_2|-|\mu_1|}f_{\mu_1}\overline{f}_{\mu_2}
||\lambda_1||^{-2}\overline{||\lambda_2||}^{-2}
\left[{\cal P}_{\mu_1,\lambda_1}^{\rm Hopf}\right]_0\cdot
\left[\overline{{\cal P}}_{\mu_2,\lambda_2}^{\rm Hopf}\right]_0
\ee

It also works similarly in the limit of $A\to\infty$: only ${\bf Z}$ is associated with the product of Hopf hyperpolynomials in this limit, (\ref{Zi}), while, say, ${\bf Z'}$ (\ref{Zpi}) does only after changing the basis: defining
\be
Z'_{str}(U_1,U_2,U_3,U_4)=\sum_{\mu_1,\mu_2,\lambda_1,\lambda_2}
\widehat{\bf Z'}\left[\begin{array}{c|c}\mu_1\,\lambda_1&\!\! \\\!\! &\mu_2\,\lambda_2\end{array}\right]M_{\mu_1}\Big\{\xi^{(1)}\Big\}
\overline{M}_{\mu_2}\Big\{\xi^{(2)}\Big\}\overline{M}_{\lambda_1^\vee}\Big\{\xi^{(3)}\Big\}
M_{\lambda_2^\vee}\Big\{\xi^{(4)}\Big\}
\ee
and re-expanding it into the basis
\be
Z'_{str}(U_1,U_2,U_3,U_4)=\sum_{\mu_1,\mu_2,\lambda_1,\lambda_2}
\widetilde{\bf Z'}\left[\begin{array}{c|c}\mu_1\,\lambda_1&\!\! \\\!\! &\mu_2\,\lambda_2\end{array}\right]M_{\mu_1}\Big\{\xi^{(1)}\Big\}
\overline{M}_{\mu_2}\Big\{\xi^{(2)}\Big\}\overline{M}_{\lambda_1^\vee}\Big\{-\xi^{(3)}\Big\}
M_{\lambda_2^\vee}\Big\{-\xi^{(4)}\Big\}
\ee
one obtains
\be
\widetilde{\bf Z'}\left[\begin{array}{c|c}\mu_1\,\lambda_1&\!\! \\\!\! &\mu_2\,\lambda_2\end{array}\right]=(A^2)^{\Sigma}\left({q\over t}\right)^{2|\mu_2|}
\left(-{q\over t}\right)^{|\lambda_1|+|\lambda_2|}f_{\lambda_1}\overline{f}_{\lambda_2}\cdot \overline{||\mu_1||}^2
\cdot ||\mu_2||^2\times\nn\\
\nn\\
\times\left[A^{|\mu_1|+|\lambda_2|}{\cal P}_{\lambda_2,\mu_1^\vee}^{\rm Hopf}(A,t^{-1},q^{-1})\right]_0\cdot
\left[A^{|\mu_2|+|\lambda_1|}{\cal P}_{\lambda_1,\mu_2^\vee}^{\rm Hopf}(A,q^{-1},t^{-1})\right]_0
\ee

\subsection{Topological 4-point function and the Hopf hyperpolynomial in the generic case}

Note that the topological 4-point functions has to be {\bf covariant} under changing the preferred directions because of the spectral duality of the DIM algebra, \cite{Miki,specdu}. This means that the properly defined partition functions should be {\bf invariant}, and have to be obtained from the same partition function upon using different bases of symmetric functions. As a particular corollary, they have to coincide at the first level, since, at this level, there is only one symmetric function. This is, indeed, the case: (\ref{3Z}).

Let us look at the next non-trivial level. Then, introducing the normalized 4-point function $\widehat{\bf Z}$, we immediately realize that
\be\label{cb}
-\sum_{\mu_1,\mu_2}(A^2)^{\Sigma}
f_{\mu_1^\vee}^2\overline{f}_{\mu_2^\vee}^2\left({q^2\over t^2}\right)^{|\mu_1|+1}
{\overline{h}_{\mu_1}\over h_{\mu_1^\vee}}
{h_{\mu_2}\over\overline{h}_{\mu_2}}\cdot
\widehat{\bf Z}\left[\begin{array}{c|c}\!\! \mu_1& [1]\cr
[1]& \mu_2\end{array}\right](A^{-1},t,q)\cdot\overline{M}_{\mu_1}\{\xi^{(1)}\}M_{\mu_2}\{\xi^{(2)}\}=\nn\\
=\sum_{\lambda_1,\lambda_2}\widehat{\bf Z''}\left[\begin{array}{c|c}\!\! [1] & \lambda_1 \\
\lambda_2& [1]\end{array}\right](A,q,t)\overline{M}_{\lambda_1}\{-\xi^{(1)}\}M_{\lambda_2}\{-\xi^{(2)}\}
\ee

In the general situation, one has to consider the three functions (using (\ref{id}) and rescaling $\xi^{(i)}$, one can definitely remove all transpositions of the Young diagrams from the basis vectors in these formulas)
\be\label{pf}
Z_{str}(U_1,U_2,U_3,U_4)=\sum_{\mu_1,\mu_2,\lambda_1,\lambda_2}
\widehat{\bf Z}\left[\begin{array}{c|c}\!\! \mu_1& \lambda_2\\
\lambda_1& \mu_2\end{array}\right]\cdot\overline{M}_{\mu_1^\vee}\Big\{\xi^{(1)}\Big\}
M_{\mu_2^\vee}\Big\{\xi^{(2)}\Big\}M_{\lambda_1}\Big\{\xi^{(3)}\Big\}
\overline{M}_{\lambda_2}\Big\{\xi^{(4)}\Big\}\nn\\
Z'_{str}(U_1,U_2,U_3,U_4)=\sum_{\mu_1,\mu_2,\lambda_1,\lambda_2}
\widehat{\bf Z'}\left[\begin{array}{c|c}\mu_1\,\lambda_1&\!\! \\\!\! &\mu_2\,\lambda_2\end{array}\right]\cdot M_{\mu_1}\Big\{\xi^{(1)}\Big\}
\overline{M}_{\mu_2}\Big\{\xi^{(2)}\Big\}\overline{M}_{\lambda_1^\vee}\Big\{\xi^{(3)}\Big\}
M_{\lambda_2^\vee}\Big\{\xi^{(4)}\Big\}\nn\\
Z''_{str}(U_1,U_2,U_3,U_4)=\sum_{\lambda_1,\lambda_2,\lambda_3,\lambda_4}
\widehat{\bf Z''}\left[\begin{array}{c|c}\!\! \lambda_3 & \lambda_4 \\
\lambda_1& \lambda_2\end{array}\right]\cdot M_{\lambda_3}\Big\{\xi^{(1)}\Big\}
\overline{M}_{\lambda_2}\Big\{\xi^{(2)}\Big\}\overline{M}_{\lambda_1^\vee}\Big\{\xi^{(3)}\Big\}
M_{\lambda_4^\vee}\Big\{\xi^{(4)}\Big\}
\ee
and one may expect that there exists an invariant partition function of four open strings given by a unique function
${\cal Z}(U_1,U_2,U_3,U_4)$ so that, in order to obtain the 4-point functions  (\ref{4p}),  (\ref{4pmu}) and (\ref{4plambda}), one has to expand it into the corresponding bases with proper changing of variables. As soon as at the level 1 there is only one symmetric function $p_1=\sum_i x_i$, i.e. the unique element of the basis, all the functions ${\bf Z}$, ${\bf Z'}$ and ${\bf Z''}$ should coincide in this case. We observed it in (\ref{3Z}).

A specific choosing the (graded) basis, however, can not allow one to associate the topological 4-point function with the Hopf hyperpolynomial in the composite representation, since they differ already at the first level: see sec.\ref{firlev}, where there is only one symmetric function.

\section{Summary}

In this paper we analyzed the relation between the 4-point functions in
the DIM based network model and the Hopf hyperpolynomial.
This is an attempt to extend our result (\ref{eq})
from \cite{L8n8} in the non-refined case, $q=t$,
when the 4-point function coincides with the HOMPLY-PT invariant for the Hopf link colored with
two composite representations, which, in turn, is the leading term
of the expansion for the HOMFLY-PT invariant for the
non-torus 4-component link $L_{8n8}$.
In the present (refined) case, the situation is much more involved.
There was a general expectation that the correspondence should lift
to the level of hyperpolynomials and the refined topological vertices.
We demonstrated that the things are not so simple.

One of the problem is that, in the refined case, there is no distinguished basis
in which the TV has an explicit cyclic symmetry,
and, in any basis, there are three non-coincident 4-point correlators though related by the spectral duality \cite{Miki,specdu}.
At the same time, the Hopf invariants look distinguished and should
be associated with some distinguished basis.
Moreover, the 4-point correlator ${\bf Z''}$ with all external legs of the same type
(``horizontal"), which one could hope to have a better symmetry property,
is the most difficult to calculate.

Below we summarize various properties of the 4-point functions, their interrelations and relations with the Hopf hyperpolynomial in a brief and pictorial way. In the figures, we draw the 4-point functions in the DIM picture \cite{AFS,AKM4OZ}, where all non-vertical lines are depicted horizontal with an assigned central charge (here is either (1,0) for the Fock representation ${\cal F}^{(1,0)}$ or (1,1) for the Fock representation ${\cal F}^{(1,1)}$). Explicit expressions for the arguments of Macdonald polynomials in the first figure can be read off from formulas (\ref{Z}) and (\ref{Zp}).

\begin{picture}(300,370)(-250,-320)

\put(-290,-2){\mbox{
$\widehat{\bf Z}\left[\begin{array}{c|c}\!\! \mu_1& \lambda_2\\
\lambda_1& \mu_2\end{array}\right]=$
}}
\put(-120,0){\vector(1,0){30}} \put(-210,5){\mbox{$\lambda_1$}}
\put(-180,0){\vector(1,0){60}}  \put(-100,5){\mbox{$\lambda_2$}}
\put(-210,0){\vector(1,0){30}}
\put(-120,-30){\vector(0,1){30}} \put(-117,-28){\mbox{$\mu_2$}}
\put(-180,0){\vector(0,1){30}}  \put(-175,25){\mbox{$\mu_1$}}
\put(-80,-3){\mbox{{  $
\sim \ D_{(\mu_1,\mu_2)}\ \displaystyle{\sum_\sigma(-)^\sigma}
\cdot \bar M_{\lambda_1^\vee/\sigma}\Big\{\bar P^{(\mu_2,\mu_1^\vee)}_{Aq/t}\Big\}
\cdot M_{\lambda_2^\vee/\sigma^\vee}\Big\{P^{(\mu_1,\mu_2^\vee)}_A\Big\}
$}}}

\put(0,-120){
\put(-290,-2){\mbox{$\widehat{\bf Z'}\left[\begin{array}{c|c}\mu_1\,\lambda_1&\!\! \\\!\! &\mu_2\,\lambda_2\end{array}\right]
 = $}}
\put(-95,0){\vector(1,0){30}} \put(-68,5){\mbox{$\lambda_1$}}
\put(-155,0){\vector(1,0){60}}
\put(-185,0){\vector(1,0){30}}  \put(-182,5){\mbox{$\lambda_2$}}
\put(-95,0){\vector(0,1){30}} \put(-150,-28){\mbox{$\mu_2$}}
\put(-155,-30){\vector(0,1){30}}  \put(-92,25){\mbox{$\mu_1$}}
\put(-55,-3){\mbox{{  $
\sim \ D_{(\mu_2^\vee,\mu_1^\vee)}\ \displaystyle{\sum_\sigma}(-)^\sigma
\cdot  M_{\lambda_1 /\sigma}\Big\{  P^{(\mu_1,\mu_2^\vee)}_{t/Aq}\Big\}
\cdot \bar M_{\lambda_2 /\sigma^\vee}\Big\{\bar P^{(\mu_2,\mu_1^\vee)}_{1/A}\Big\}
$}}}
}

\put(120,-60){
\put(0,0){\vector(0,1){40}}\put(0,0){\vector(0,-1){40}}
\put(10,-30){\line(0,1){60}} \put(10,-30){\line(1,0){60}}
\put(70,-30){\line(0,1){60}} \put(10,30){\line(1,0){60}}
\put(15,20){\mbox{$q\longleftrightarrow t$}}
\put(15,5){\mbox{$A\longleftrightarrow A^{-1}$}}
\put(15,-10){\mbox{$\lambda_i \longleftrightarrow \lambda_i^\vee$ }}
\put(15,-25){\mbox{$\mu_1 \longleftrightarrow \mu_2$}}
}

\put(-20,-210){
\put(-270,-25){\mbox{$\widehat{\bf Z''}\left[\begin{array}{c|c}\!\! \lambda_3 & \lambda_4 \\
\lambda_1& \lambda_2\end{array}\right]
= $}}
\put(-140,0){\vector(1,0){40}} \put(-105,5){\mbox{$\lambda_4$}}
\put(-180,0){\vector(1,0){40}} \put(-178,5){\mbox{$\lambda_2$}}
\put(-140,-50){\vector(0,1){50}}
\put(-140,-50){\vector(1,0){40}}  \put(-105,-45){\mbox{$\lambda_3$}}
\put(-180,-50){\vector(1,0){40}}  \put(-178,-45){\mbox{$\lambda_1$}}
}

\put(0,-180){
\put(0,0){\vector(0,1){40}}\put(0,0){\vector(0,-1){40}}
\put(10,-30){\line(0,1){60}} \put(10,-30){\line(1,0){220}}
\put(230,-30){\line(0,1){60}} \put(10,30){\line(1,0){220}}
{\footnotesize
\put(15,10){\mbox{$
\ \displaystyle{\sum_{\lambda_3,\lambda_4}}\ 
\widehat{\bf Z''}\left[\begin{array}{c|c}\!\! [1] & \lambda_1 \\
\lambda_2& [1]\end{array}\right]\overline{M}_{\lambda_1}\{-\xi\}M_{\lambda_2}\{-\tilde\xi\}
\sim
$}}
\put(15,-15){\mbox{$
\ \sim \displaystyle{\sum_{\mu_1,\mu_2}}\ 
{\cal C}^{\mu_1,\mu_2}\widehat{\bf Z'}\left[\begin{array}{c|c}\mu_2\,[1]&\!\! \\\!\! &\mu_1\,[1]\end{array}\right]\overline{M}_{\mu_1}\{\xi\}M_{\mu_2}\{\tilde\xi\}
$}}
}
}

\linethickness{0.5mm}
\put(-280,-70){\line(1,0){370}}
\put(-280,-190){\line(1,0){250}}

\end{picture}

\vspace{-1.5cm}

Remarkably, the two functions ${\bf Z}$ and ${\bf Z'}$ are related by the transformation in the box, see (\ref{flop}).
This connection is related to the flop operation discussed in \cite{AK08} with an additional transformation, (\ref{12}). This relation is {\it not} an apparent symmetries of the network diagrams
(reversion of arrows for $\lambda$'s and $\mu$ has clear interpretation
in this transformation,
but interchange of $\mu_1$ and $\mu_2$ is not).
This newly emerging after summation over intermediate states
{\bf effective symmetry} is a manifestation of the $SL(2,\mathbb{Z})$-duality of the DIM algebra \cite{Miki}.

Another one relates
this symmetry-related pair ${\bf Z}\cong {\bf Z'}$ and ${\bf Z''}$.
At the DIM level, these are essentially different quantities, still they are related by the spectral duality and, at the technical level,
by a "change of basis" like
\be
\sum_{\lambda_3,\lambda_4}
\widehat{\bf Z''}\left[\begin{array}{c|c}\!\! [1] & \lambda_1 \\
\lambda_2& [1]\end{array}\right]\overline{M}_{\lambda_1}\{-\xi^{(1)}\}M_{\lambda_2}\{-\xi^{(2)}\}
\sim \sum_{\mu_1,\mu_2} {\cal C}^{\mu_1,\mu_2}(q,t)\cdot
\widehat{\bf Z'}\left[\begin{array}{c|c}\mu_2\,[1]&\!\! \\\!\! &\mu_1\,[1]\end{array}\right]\overline{M}_{\mu_1}\{\xi^{(1)}\}M_{\mu_2}\{\xi^{(2)}\}\nn
\ee
with some coefficients ${\cal C}^{\mu_1,\mu_2}(q,t)$.
For the diagrams $\lambda_3$ and $\lambda_4$ larger than $[1]$, one has to change the basis of all four Macdonald polynomials, and what remains is just a single linear relation between the functions at the l.h.s and r.h.s. unless one allows the coefficients ${\cal C}^{\mu_1,\mu_2}$ depend also on $A$.

Thus, the choice of a proper basis looks very essential for establishing any relations between the 4-point functions and the Hopf invariants.
However, we demonstrated that the Hopf hyperpolynomial in composite representations does not coincide with the refined 4-point function in any basis: first of all, it realizes a chirality property, that is, these Hopf hyperpolynomials are "chiral" bilinears
of Macdonald functions $M\cdot M$, while convolutions of the topological vertices
involve non-chiral combinations $M\cdot \overline{M}$. This is illustrated with the figure, where the bold arrow means an additional changing the basis (see sec.7.2) and underlining the Hopf hyperpolynomials means the change $(q,t)\to(q^{-1},t^{-1})$. Note that the substitution $(q,t)\to(t^{-1},q^{-1})$ is equivalent, up to a factor, to transposition of all Young diagrams.

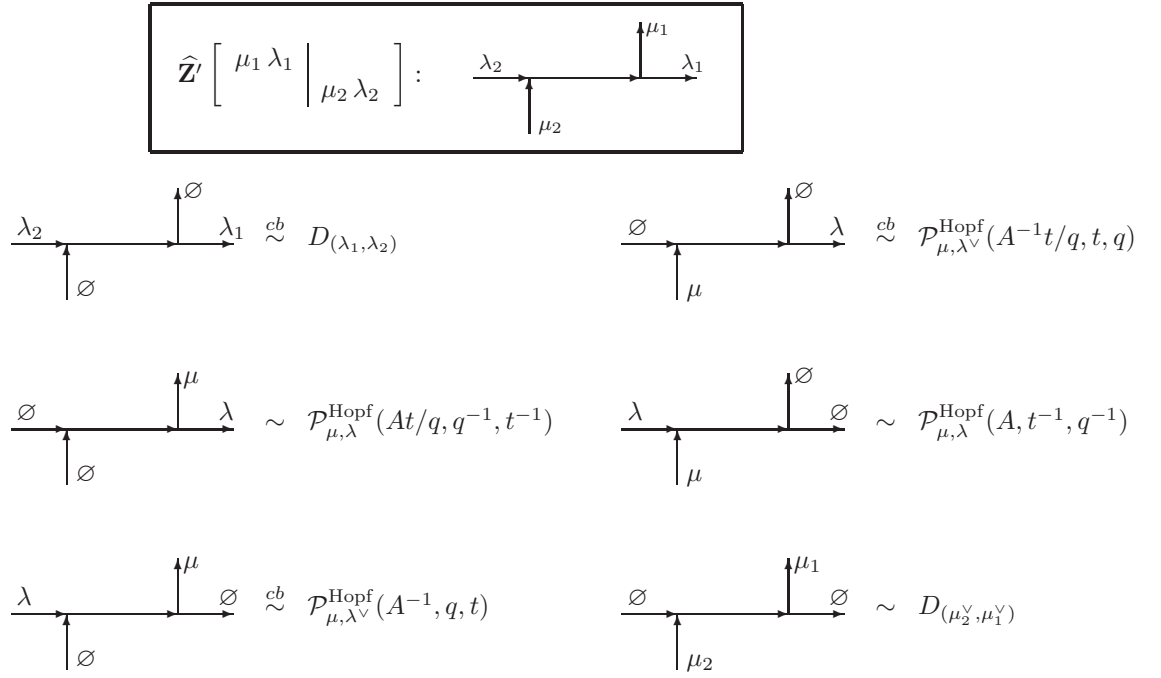
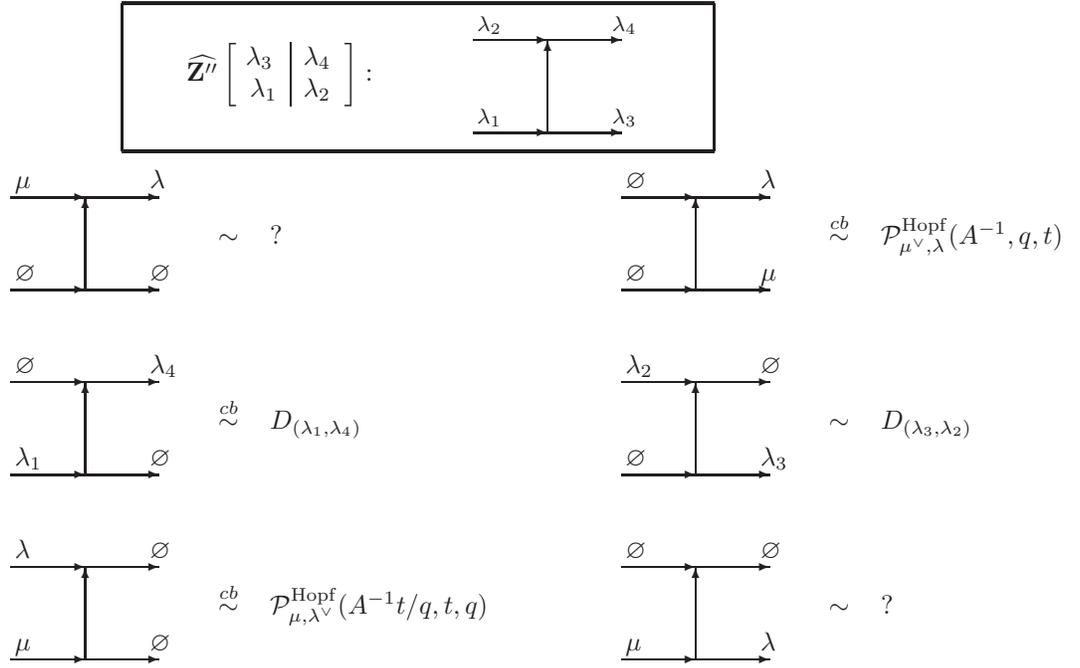
\begin{picture}(300,170)(-230,-120)
\unitlength=.7pt
\put(-390,-2){\mbox{
$\widehat{\bf Z}\left[\begin{array}{c|c}\!\! \mu_1& \lambda_2\\
\lambda_1& \mu_2\end{array}\right]=$
}}
\put(-170,0){\vector(1,0){30}} \put(-260,5){\mbox{$\lambda_1$}}
\put(-230,0){\vector(1,0){60}}  \put(-150,5){\mbox{$\lambda_2$}}
\put(-260,0){\vector(1,0){30}}
\put(-170,-30){\vector(0,1){30}} \put(-167,-28){\mbox{$\mu_2$}}
\put(-230,0){\vector(0,1){30}}  \put(-225,25){\mbox{$\mu_1$}}
\put(-120,5){\vector(3,1){50}}\put(-120,5.5){\line(3,1){45}}\put(-120,4.5){\line(3,1){45}}
\put(-60,18){\mbox{$\left[{\cal P}_{\mu_1,\lambda_1}^{\rm Hopf}\right]_0\cdot
\left[\overline{{\cal P}}_{\mu_2,\lambda_2}^{\rm Hopf}\right]_0$}}
\put(-120,20){\mbox{\footnotesize{$A=0$}}}
\put(-120,-5){\vector(3,-1){50}}\put(-60,-25){\mbox{$\left[\overline{\underline{\cal P}}_{\lambda_1^\vee,\mu_2}^{\rm Hopf}\right]_0\cdot
\left[\underline{\cal P}_{\lambda_2^\vee,\mu_1}^{\rm Hopf}\right]_0$}}
\put(-120,-25){\mbox{\footnotesize{$A=\infty$}}}

\put(300,5){\vector(-3,1){50}}
\put(110,18){\mbox{$\left[{\cal P}_{\lambda_1,\mu_1}^{\rm Hopf}\right]_0\cdot
\left[{\cal P}_{\lambda_2,\mu_2}^{\rm Hopf}\right]_0$}}
\put(275,20){\mbox{\footnotesize{$A=0$}}}
\put(300,-5){\vector(-3,-1){50}}\put(110,-25){\mbox{$\left[\underline{\cal P}_{\lambda_1,\mu_2}^{\rm Hopf}\right]_0\cdot
\left[\underline{\cal P}_{\lambda_2,\mu_1}^{\rm Hopf}\right]_0$}}
\put(275,-25){\mbox{\footnotesize{$A=\infty$}}}
 \put(310,-5){\mbox{${\cal P}_{(\mu_1,\mu_2),(\lambda_1,\lambda_2)}^{\rm Hopf}$}}

\put(-10,-120){
\put(-390,-2){\mbox{$\widehat{\bf Z'}\left[\begin{array}{c|c}\mu_1\,\lambda_1&\!\! \\\!\! &\mu_2\,\lambda_2\end{array}\right]
 = $}}
\put(-155,0){\vector(1,0){30}} \put(-133,5){\mbox{$\lambda_1$}}
\put(-215,0){\vector(1,0){60}}
\put(-245,0){\vector(1,0){30}}  \put(-242,5){\mbox{$\lambda_2$}}
\put(-155,0){\vector(0,1){30}} \put(-210,-28){\mbox{$\mu_2$}}
\put(-215,-30){\vector(0,1){30}}  \put(-152,25){\mbox{$\mu_1$}}
}

\put(-120,-115){\vector(3,1){50}}
\put(-60,-102){\mbox{$\left[\underline{\cal P}_{\lambda_1,\mu_1}^{\rm Hopf}\right]_0\cdot
\left[\overline{\underline{\cal P}}_{\lambda_2,\mu_2}^{\rm Hopf}\right]_0$}}
\put(-120,-100){\mbox{\footnotesize{$A=0$}}}
\put(-120,-125){\vector(3,-1){50}}\put(-120,-124.5){\line(3,-1){45}}\put(-120,-125.5){\line(3,-1){45}}
\put(-60,-145){\mbox{$\left[\overline{\cal P}_{\lambda_1^\vee,\mu_2}^{\rm Hopf}\right]_0\cdot
\left[{\cal P}_{\lambda_2^\vee,\mu_1}^{\rm Hopf}\right]_0$}}
\put(-120,-145){\mbox{\footnotesize{$A=\infty$}}}

\put(300,-115){\vector(-3,1){50}}
\put(110,-102){\mbox{$\left[{\cal P}_{\lambda_1,\mu_1}^{\rm Hopf}\right]_0\cdot
\left[{\cal P}_{\lambda_2,\mu_2}^{\rm Hopf}\right]_0$}}
\put(275,-100){\mbox{\footnotesize{$A=0$}}}
\put(300,-125){\vector(-3,-1){50}}\put(110,-145){\mbox{$\left[\underline{\cal P}_{\lambda_1,\mu_2}^{\rm Hopf}\right]_0\cdot
\left[\underline{\cal P}_{\lambda_2,\mu_1}^{\rm Hopf}\right]_0$}}
\put(275,-145){\mbox{\footnotesize{$A=\infty$}}}
 \put(310,-125){\mbox{${\cal P}_{(\mu_1,\mu_2),(\lambda_1,\lambda_2)}^{\rm Hopf}$}}

\put(90,-150){\line(0,1){200}}

\end{picture}

\bigskip

\noindent
We can observe  that, in the $A=0$ limit, the answers both for the 4-point functions and for the Hopf hyperpolynomial in the composite representation are factorized into the entries that depend on the pairs of Young diagrams associated with the first and second vertices accordingly. Moreover, all these entries are the corresponding Hopf invariant asymptotics, up to possible permutation of $q$ and $t$. At the same time, in the $A=\infty$ limit, the structure of answers remains the same, the pairs of the Young diagram being associated with the first and second vertices {\it after the conifold transition}, as it should be. Hence, both for the 4-point functions and for the Hopf hyperpolynomial in the composite representation have similar ``initial" conditions at $A=0$ (up to possible permutation of $q$ and $t$) and respect the conifold transition, still, the continuation to non-zero $A$ seems to be different.

Thus, the situation is even worse than the different chiral structures of the two quantities: the example of lower rank representations demonstrates that the difference is much deeper, see sec.\ref{firlev}. Another drawback of the refined 4-point functions is that they are {\bf not symmetric} under the permutation $(\mu_1,\mu_2)\leftrightarrow (\lambda_1,\lambda_2)$, which is the case for the Hopf link invariant, since it just interchanges the Hopf link components.

At the same time, the 4-point functions with only two non-trivial diagrams are often associated either with the Hopf hyperpolynomials, or with the Macdonald dimensions. In order to illustrate this better, we collect together all such cases here. Here {\it cb} means an additional changing the basis.

\begin{picture}(300,290)(-230,-185)
\unitlength=.7pt

{\linethickness{0.3mm}
\put(-170,130){\line(1,0){320}}
\put(-170,50){\line(1,0){320}}
\put(-170,50){\line(0,1){80}}
\put(150,50){\line(0,1){80}}
}

\put(-140,88){\mbox{
$\widehat{\bf Z}\left[\begin{array}{c|c}\!\! \mu_1& \lambda_2\\
\lambda_1& \mu_2\end{array}\right]:$
}}

\put(250,90){
\footnotesize{
\put(-170,0){\vector(1,0){30}} \put(-260,5){\mbox{$\lambda_1$}}
\put(-230,0){\vector(1,0){60}}  \put(-150,5){\mbox{$\lambda_2$}}
\put(-260,0){\vector(1,0){30}}
\put(-170,-30){\vector(0,1){30}} \put(-167,-28){\mbox{$\mu_2$}}
\put(-230,0){\vector(0,1){30}}  \put(-225,25){\mbox{$\mu_1$}}}
}

\put(0,0){
\put(-170,0){\vector(1,0){30}} \put(-260,5){\mbox{$\lambda_1$}}
\put(-230,0){\vector(1,0){60}}  \put(-150,5){\mbox{$\lambda_2$}}
\put(-260,0){\vector(1,0){30}}
\put(-170,-30){\vector(0,1){30}} \put(-167,-28){\mbox{$\varnothing$}}
\put(-230,0){\vector(0,1){30}}  \put(-225,25){\mbox{$\varnothing$}}
\put(-118,0){\mbox{$\stackrel{cb}{\sim}$}}
\put(-90,0){\mbox{$D_{(\lambda_1,\lambda_2)}$}}
}

\put(330,0){
\put(-170,0){\vector(1,0){30}} \put(-260,5){\mbox{$\lambda$}}
\put(-230,0){\vector(1,0){60}}  \put(-150,5){\mbox{$\varnothing$}}
\put(-260,0){\vector(1,0){30}}
\put(-170,-30){\vector(0,1){30}} \put(-167,-28){\mbox{$\varnothing$}}
\put(-230,0){\vector(0,1){30}}  \put(-225,25){\mbox{$\mu$}}
\put(-118,0){\mbox{$\stackrel{cb}{\sim}$}}
\put(-90,0){\mbox{${\cal P}_{\mu,\lambda}^{\rm Hopf}(At/q,q,t)$}}
}

\put(0,-100){
\put(-170,0){\vector(1,0){30}} \put(-260,5){\mbox{$\lambda$}}
\put(-230,0){\vector(1,0){60}}  \put(-150,5){\mbox{$\varnothing$}}
\put(-260,0){\vector(1,0){30}}
\put(-170,-30){\vector(0,1){30}} \put(-167,-28){\mbox{$\mu$}}
\put(-230,0){\vector(0,1){30}}  \put(-225,25){\mbox{$\varnothing$}}
\put(-118,0){\mbox{$\sim$}}
\put(-90,0){\mbox{$\displaystyle{{\cal P}_{\mu,\lambda^\vee}^{\rm Hopf}\Big(A^{-1}{t\over q},{1\over t},{1\over q}\Big)}$}}
}

\put(330,-100){
\put(-170,0){\vector(1,0){30}} \put(-260,5){\mbox{$\varnothing$}}
\put(-230,0){\vector(1,0){60}}  \put(-150,5){\mbox{$\lambda$}}
\put(-260,0){\vector(1,0){30}}
\put(-170,-30){\vector(0,1){30}} \put(-167,-28){\mbox{$\varnothing$}}
\put(-230,0){\vector(0,1){30}}  \put(-225,25){\mbox{$\mu$}}
\put(-118,0){\mbox{$\sim$}}
\put(-90,0){\mbox{${\cal P}_{\mu,\lambda^\vee}^{\rm Hopf}(A^{-1},q^{-1},t^{-1})$}}
}

\put(0,-200){
\put(-170,0){\vector(1,0){30}} \put(-260,5){\mbox{$\varnothing$}}
\put(-230,0){\vector(1,0){60}}  \put(-150,5){\mbox{$\lambda$}}
\put(-260,0){\vector(1,0){30}}
\put(-170,-30){\vector(0,1){30}} \put(-167,-28){\mbox{$\mu$}}
\put(-230,0){\vector(0,1){30}}  \put(-225,25){\mbox{$\varnothing$}}
\put(-118,0){\mbox{$\stackrel{cb}{\sim}$}}
\put(-90,0){\mbox{${\cal P}_{\mu,\lambda}^{\rm Hopf}(A,t,q)$}}
}

\put(330,-200){
\put(-170,0){\vector(1,0){30}} \put(-260,5){\mbox{$\varnothing$}}
\put(-230,0){\vector(1,0){60}}  \put(-150,5){\mbox{$\varnothing$}}
\put(-260,0){\vector(1,0){30}}
\put(-170,-30){\vector(0,1){30}} \put(-167,-28){\mbox{$\mu_2$}}
\put(-230,0){\vector(0,1){30}}  \put(-225,25){\mbox{$\mu_1$}}
\put(-118,0){\mbox{$\sim$}}
\put(-90,0){\mbox{$D_{(\mu_1,\mu_2)}$}}
}

\end{picture}

\begin{picture}(300,270)(-230,-185)
\unitlength=.7pt
\put(-160,88){\mbox{
$\widehat{\bf Z'}\left[\begin{array}{c|c}\mu_1\,\lambda_1&\!\! \\\!\! &\mu_2\,\lambda_2\end{array}\right]:$
}}

{\linethickness{0.3mm}
\put(-170,130){\line(1,0){320}}
\put(-170,50){\line(1,0){320}}
\put(-170,50){\line(0,1){80}}
\put(150,50){\line(0,1){80}}
}

\put(250,90){
\footnotesize{
\put(-155,0){\vector(1,0){30}} \put(-133,5){\mbox{$\lambda_1$}}
\put(-215,0){\vector(1,0){60}}
\put(-245,0){\vector(1,0){30}}  \put(-242,5){\mbox{$\lambda_2$}}
\put(-155,0){\vector(0,1){30}} \put(-210,-28){\mbox{$\mu_2$}}
\put(-215,-30){\vector(0,1){30}}  \put(-152,25){\mbox{$\mu_1$}}
}}

\put(0,0){
\put(-155,0){\vector(1,0){30}} \put(-133,5){\mbox{$\lambda_1$}}
\put(-215,0){\vector(1,0){60}}
\put(-245,0){\vector(1,0){30}}  \put(-242,5){\mbox{$\lambda_2$}}
\put(-155,0){\vector(0,1){30}} \put(-210,-28){\mbox{$\varnothing$}}
\put(-215,-30){\vector(0,1){30}}  \put(-152,25){\mbox{$\varnothing$}}
\put(-108,0){\mbox{$\stackrel{cb}{\sim}$}}
\put(-85,0){\mbox{$D_{(\lambda_1,\lambda_2)}$}}
}

\put(330,0){
\put(-155,0){\vector(1,0){30}} \put(-133,5){\mbox{$\lambda$}}
\put(-215,0){\vector(1,0){60}}
\put(-245,0){\vector(1,0){30}}  \put(-242,5){\mbox{$\varnothing$}}
\put(-155,0){\vector(0,1){30}} \put(-210,-28){\mbox{$\mu$}}
\put(-215,-30){\vector(0,1){30}}  \put(-152,25){\mbox{$\varnothing$}}
\put(-108,0){\mbox{$\stackrel{cb}{\sim}$}}
\put(-85,0){\mbox{${\cal P}_{\mu,\lambda^\vee}^{\rm Hopf}(A^{-1}t/q,t,q)$}}
}

\put(0,-100){
\put(-155,0){\vector(1,0){30}} \put(-133,5){\mbox{$\lambda$}}
\put(-215,0){\vector(1,0){60}}
\put(-245,0){\vector(1,0){30}}  \put(-242,5){\mbox{$\varnothing$}}
\put(-155,0){\vector(0,1){30}} \put(-210,-28){\mbox{$\varnothing$}}
\put(-215,-30){\vector(0,1){30}}  \put(-152,25){\mbox{$\mu$}}
\put(-108,0){\mbox{${\sim}$}}
\put(-85,0){\mbox{${\cal P}_{\mu,\lambda}^{\rm Hopf}(At/q,q^{-1},t^{-1})$}}
}

\put(330,-100){
\put(-155,0){\vector(1,0){30}} \put(-133,5){\mbox{$\varnothing$}}
\put(-215,0){\vector(1,0){60}}
\put(-245,0){\vector(1,0){30}}  \put(-242,5){\mbox{$\lambda$}}
\put(-155,0){\vector(0,1){30}} \put(-210,-28){\mbox{$\mu$}}
\put(-215,-30){\vector(0,1){30}}  \put(-152,25){\mbox{$\varnothing$}}
\put(-108,0){\mbox{${\sim}$}}
\put(-85,0){\mbox{${\cal P}_{\mu,\lambda}^{\rm Hopf}(A,t^{-1},q^{-1})$}}
}

\put(0,-200){
\put(-155,0){\vector(1,0){30}} \put(-133,5){\mbox{$\varnothing$}}
\put(-215,0){\vector(1,0){60}}
\put(-245,0){\vector(1,0){30}}  \put(-242,5){\mbox{$\lambda$}}
\put(-155,0){\vector(0,1){30}} \put(-210,-28){\mbox{$\varnothing$}}
\put(-215,-30){\vector(0,1){30}}  \put(-152,25){\mbox{$\mu$}}
\put(-108,0){\mbox{$\stackrel{cb}{\sim}$}}
\put(-85,0){\mbox{${\cal P}_{\mu,\lambda^\vee}^{\rm Hopf}(A^{-1},q,t)$}}
}

\put(330,-200){
\put(-155,0){\vector(1,0){30}} \put(-133,5){\mbox{$\varnothing$}}
\put(-215,0){\vector(1,0){60}}
\put(-245,0){\vector(1,0){30}}  \put(-242,5){\mbox{$\varnothing$}}
\put(-155,0){\vector(0,1){30}} \put(-210,-28){\mbox{$\mu_2$}}
\put(-215,-30){\vector(0,1){30}}  \put(-152,25){\mbox{$\mu_1$}}
\put(-108,0){\mbox{${\sim}$}}
\put(-85,0){\mbox{$D_{(\mu_2^\vee,\mu_1^\vee)}$}}
}

\end{picture}

\begin{picture}(300,280)(-230,-185)
\unitlength=.7pt

{\linethickness{0.3mm}
\put(-170,130){\line(1,0){320}}
\put(-170,50){\line(1,0){320}}
\put(-170,50){\line(0,1){80}}
\put(150,50){\line(0,1){80}}
}

\put(-140,88){\mbox{
$\widehat{\bf Z''}\left[\begin{array}{c|c}\!\! \lambda_3 & \lambda_4 \\
\lambda_1& \lambda_2\end{array}\right]:$
}}

\put(250,110){
\footnotesize{
\put(-190,0){\vector(1,0){40}} \put(-155,5){\mbox{$\lambda_4$}}
\put(-230,0){\vector(1,0){40}} \put(-228,5){\mbox{$\lambda_2$}}
\put(-190,-50){\vector(0,1){50}}
\put(-190,-50){\vector(1,0){40}}  \put(-155,-45){\mbox{$\lambda_3$}}
\put(-230,-50){\vector(1,0){40}}  \put(-228,-45){\mbox{$\lambda_1$}}
}}

\put(0,0){
\put(-190,25){\vector(1,0){40}} \put(-155,30){\mbox{$\lambda$}}
\put(-230,25){\vector(1,0){40}} \put(-228,30){\mbox{$\mu$}}
\put(-190,-25){\vector(0,1){50}}
\put(-190,-25){\vector(1,0){40}}  \put(-155,-20){\mbox{$\varnothing$}}
\put(-230,-25){\vector(1,0){40}}  \put(-228,-20){\mbox{$\varnothing$}}
\put(-118,0){\mbox{${\sim}$}}
\put(-90,0){\mbox{$?$}}
}

\put(330,0){
\put(-190,25){\vector(1,0){40}} \put(-155,30){\mbox{$\lambda$}}
\put(-230,25){\vector(1,0){40}} \put(-228,30){\mbox{$\varnothing$}}
\put(-190,-25){\vector(0,1){50}}
\put(-190,-25){\vector(1,0){40}}  \put(-155,-20){\mbox{$\mu$}}
\put(-230,-25){\vector(1,0){40}}  \put(-228,-20){\mbox{$\varnothing$}}
\put(-118,0){\mbox{$\stackrel{cb}{\sim}$}}
\put(-90,0){\mbox{${\cal P}_{\mu^\vee,\lambda}^{\rm Hopf}(A^{-1},q,t)$}}
}

\put(0,-100){
\put(-190,25){\vector(1,0){40}} \put(-155,30){\mbox{$\lambda_4$}}
\put(-230,25){\vector(1,0){40}} \put(-228,30){\mbox{$\varnothing$}}
\put(-190,-25){\vector(0,1){50}}
\put(-190,-25){\vector(1,0){40}}  \put(-155,-20){\mbox{$\varnothing$}}
\put(-230,-25){\vector(1,0){40}}  \put(-228,-20){\mbox{$\lambda_1$}}
\put(-118,0){\mbox{$\stackrel{cb}{\sim}$}}
\put(-90,0){\mbox{$D_{(\lambda_1,\lambda_4)}$}}
}

\put(330,-100){
\put(-190,25){\vector(1,0){40}} \put(-155,30){\mbox{$\varnothing$}}
\put(-230,25){\vector(1,0){40}} \put(-228,30){\mbox{$\lambda_2$}}
\put(-190,-25){\vector(0,1){50}}
\put(-190,-25){\vector(1,0){40}}  \put(-155,-20){\mbox{$\lambda_3$}}
\put(-230,-25){\vector(1,0){40}}  \put(-228,-20){\mbox{$\varnothing$}}
\put(-118,0){\mbox{$\sim$}}
\put(-90,0){\mbox{$D_{(\lambda_3,\lambda_2)}$}}
}

\put(0,-200){
\put(-190,25){\vector(1,0){40}} \put(-155,30){\mbox{$\varnothing$}}
\put(-230,25){\vector(1,0){40}} \put(-228,30){\mbox{$\lambda$}}
\put(-190,-25){\vector(0,1){50}}
\put(-190,-25){\vector(1,0){40}}  \put(-155,-20){\mbox{$\varnothing$}}
\put(-230,-25){\vector(1,0){40}}  \put(-228,-20){\mbox{$\mu$}}
\put(-118,0){\mbox{$\stackrel{cb}{\sim}$}}
\put(-90,0){\mbox{${\cal P}_{\mu,\lambda^\vee}^{\rm Hopf}(A^{-1}t/q,t,q)$}}
}

\put(330,-200){
\put(-190,25){\vector(1,0){40}} \put(-155,30){\mbox{$\varnothing$}}
\put(-230,25){\vector(1,0){40}} \put(-228,30){\mbox{$\varnothing$}}
\put(-190,-25){\vector(0,1){50}}
\put(-190,-25){\vector(1,0){40}}  \put(-155,-20){\mbox{$\lambda$}}
\put(-230,-25){\vector(1,0){40}}  \put(-228,-20){\mbox{$\mu$}}
\put(-118,0){\mbox{$\sim$}}
\put(-90,0){\mbox{$?$}}
}

\end{picture}

The clear correspondences between the Hopf hyperpolynomials and the 4-point functions with only two non-trivial diagrams probably mean that the gluing of two vertices together is inconsistent with the group properties, that is, with the composite representations. To put it differently, conjugation of the representations in the Hopf hyperpolynomial is not consistent with the topological vertex picture.

It looks so that all the problems described are intimately related to the presence of two different families
of Macdonald polynomials $M$ and $\overline{M}$
in the topological vertex.
At the unrefined (Schur) level, they just coincide, but, at the Macdonald level, they get unrelated.
The two families correspond to building the Macdonald polynomials, and, more generally, Kerov functions,
by triangular transformations of the Schur polynomials \cite{Mac,Kerov,MMKerov}, which start from two opposite ends:
$M$ are associated with the lexicographic ordering, beginning at antisymmetric representations,
while $\overline{M}$, with the inverse one, beginning at symmetric representations.
The fact that the pair $(M,\overline{M})$ is associated with the bra/ket states in the network models
is the basic one in representation theory of the DIM algebra, and it is very difficult to change.
Our claim is that it results in a drastic deviation between the network correlators
and knot hyperpolynomials.
This makes the tangle calculus program in the refined case much less naive and more challenging.
One may expect many new and surprising discoveries on this way.

What we discussed in this paper is just the bottom line of the following hypothetical table:

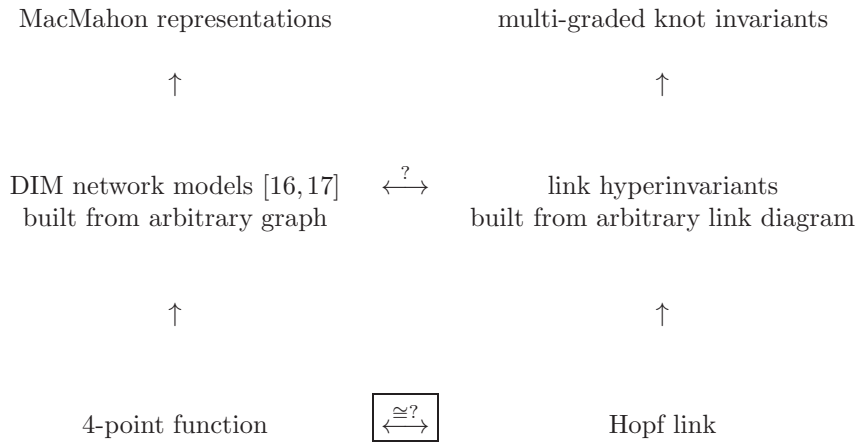
\begin{center}
\begin{tabular}{ccc}
MacMahon representations &
& multi-graded knot invariants
\\ \\
$\uparrow$ && $\uparrow$ \\
\\ \\
DIM network models \cite{network,AKM4OZ}  & $\stackrel{?}{\longleftrightarrow}$ & link hyperinvariants
\\
built from arbitrary graph & & built from arbitrary link diagram
\\ \\ \\
$\uparrow$ && $\uparrow$ \\
\\ \\
4-point function &$\boxed{\stackrel{\cong ?}{\longleftrightarrow}}$& Hopf link
\end{tabular}
\end{center}

\bigskip

\noindent
We explained certain difficulties in building the equivalence in the box.
But most important is that the both sides have their own far-going generalizations:
one can construct from the topological vertices multi-point functions
associated with arbitrary networks.
The Hopf link is just a simplest member of the family of all knots and links,
constructed from the {\it lock} and more complicated elementary tangles
as explained in \cite{tangles}.
Since the bottom line in the table identifies (currently with reservations)
the {\it lock} with the pair of topological vertices,
one can hope that generic networks will be similarly related to generic
link diagrams, though there is a long way to go in order to understand,
if this relation indeed exists and what it is.
Importantly, this relation should be ``functorial" w.r.t. the
relation between Young diagrams at the l.h.s. and representations at the r.h.s.:
the difficulties with establishing this property at the bottom line
was the main subject of the present paper:
we have not tamed them completely, but demonstrated that the hope persists.
This complements another non-trivial piece of evidence:
that hyperpolynomial and even superpolynomial
calculi actually preserve some crucial implications of the Reshetikhin-Turaev
representation theory approach \cite{MRT}, especially the {\it evolution} properties
as originally suggested  in \cite{DMMSS,evo} and now broadly checked within various
contexts \cite{Che,AENV,M16,ArthSha,KNTZ,Anokhevo}.

Furthermore, both columns have further generalizations:
to the MacMahon representations on the DIM side and to multi-graded knot invariants
on the link side.
At the moment they have nothing in common: the MacMahon representation means lifting from
the Young diagrams to the plane partitions \cite{AKMMSZmacmahon},
of the multi-gradings only the forth is currently under study \cite{GGS,diffarth,KM17fe,NawOb}.
Still, the very existence of further generalizations opens a very broad perspective
for  the study of {\it tangle calculus}, the ambitious program \cite{L8n8,tangles}
to {\bf relate link hyperpolynomials with representation theory of DIM-like algebras}.
We hope that this paper is a good illustration of what are the kind of obstacles
to overcome on this long way.

\section*{Acknowledgements}

Our work is supported in part by Grants-in-Aid for Scientific Research
(17K05275) (H.A.), (15H05738, 18K03274) (H.K.) and JSPS Bilateral Joint Projects (JSPS-RFBR collaboration)
``Elliptic algebras, vertex operators and link invariants'' from MEXT, Japan. It is also partly supported by the grant of the Foundation for the Advancement of Theoretical Physics ``BASIS" (A.Mir., A.Mor.), by  RFBR grants 19-01-00680 (A.Mir.) and 19-02-00815 (A.Mor.), by joint grants 19-51-53014-GFEN-a (A.Mir., A.Mor.), 19-51-50008-YaF-a (A.Mir.), 18-51-05015-Arm-a (A.Mir., A.Mor.), 18-51-45010-IND-a (A.Mir., A.Mor.).

\section*{Appendix}

\subsection*{A1. Macdonald polynomials}

The Macdonald polynomials
\be
M_R:=\sum_{Q}{\cal K}_{RQ}(q,t)\cdot\Sch_Q
\ee
where the Kostka-Macdonald coefficients ${\cal K}_{RQ}(q,t)$ are Laurent polynomials in $q$ and $t$ with positive integer coefficients \cite{Garcia}, and ${\cal K}_{RR}(q,t)=1$, ${\cal K}_{RQ}(q,t)=0$ at $Q>R$, with the lexicographical order of partitions. They
are defined by the orthogonality condition
\be
\Big<M_R,M_{R'}\Big>=0\ \ \ \ \ \hbox{unless }R=R'
\ee
with the scalar product induced by
\be
<p_{\Delta},p_{\Delta'}>=z_\Delta\, \delta_{\Delta,\Delta'}
\prod_i{\{q^{\delta_i}\}\over\{t^{\delta_i}\}}
\ee
where $z_\Delta$ is the order of automorphism of the Young diagram $\Delta$
and we have scaled the original inner product of Macdonald \cite{Mac} by the factor of $(q/t)^{|\Delta|}$.
These conditions unambiguously define the Macdonald polynomials. In particular, the full orthogonality relations look like
\be\label{MOR}
\Big<M_R,M_{R'}\Big>={\overline{h}_{R^\vee}\over h_R}\, \delta_{R,R'}:=||R||^2\, \delta_{R,R'}
\ee
where
\be\label{hook}
h_\nu(q,t):=\prod_{i,j\in \nu} (q^{\nu_i-j}t^{\nu'_j-i+1}-q^{j-\nu_i}t^{i-\nu'_j-1})
\ee
The dual Macdonald polynomials are defined as
\be
Q_R:={h_R\over \overline{h}_{R^\vee}}M_R
\ee

In fact, instead of the orthogonality relations, one can use a determinant formula \cite{LLM}
\be
M_R=\mathfrak{N}_R\cdot\det_{P,Q\ge R} {\cal C}^{(R)}_{PQ}
\ee
where the first row of the matrix, ${\cal C}^{(R)}_{RQ}=\Sch_Q \left\{\frac{\{t^k\}}{\{q^k\}}p_k\right\}$ and all others are:
${\cal C}^{(R)}_{PQ}=\Big[S_Q\Big]_P ^{(R)}$. Here $S_Q:=\det_{i,j}S_{Q_i-i+j}\cdot S_0^{l(R)-l(Q)}$ and $\Big[S_Q\Big]_P^{(R)} $ denotes the sum of all ordered terms (taken with signs) in $S_Q$ of the cycle type $P$, with each term $S_{i_1}\ldots S_{i_n}$, $n=|R|$ substituted by the
$F(R,[i_k]):=\sum_{k=1}^n \Big(q^{2R_i}-q^{2i_k}\Big) t^{2n-2k}$ with taking into account the signs. The normalization factor is
\be
\mathfrak{N}_R={\overline{h}_{R^\vee}\over h_R}\prod_{Q>R}{1\over {\cal C}^{(R)}_{QQ}}
\ee

For instance, consider $Q=[2,1,1,1]$. Then,
\be
S_{Q}=\left|
\begin{array}{cccc}
S_2&S_3&S_4&S_5\\
S_0&S_1&S_2&S_3\\
0&S_0&S_1&S_2\\
0&0&S_0&S_1
\end{array}
\right|\cdot S_0
=\ldots-S_2S_1S_2S_0S_0-S_2S_2S_0S_1S_0+\ldots
\ee
which, e.g., for $R=[1,1,1,1,1]$ and $P=[2,2,1]$ gives rise to
$$
{\cal C}^{([1,1,1,1,1])}_{[2,1,1],[1,1,1,1]}=\Big[S_{[1,1,1,1]}\Big]_{[2,1,1]} ^{([1,1,1,1])}
=-F([1,1,1,1,1],[2,1,2,0,0])-F([1,1,1,1,1],[2,2,0,1,0])=
$$
\vspace{-.5cm}
\be
=(q^2-1)(2q^2t^8+q^2t^6+q^2t^4-t^4-t^2-2)
\ee

\subsection*{A2. Skew Macdonald polynomials}

Like skew Schur polynomials, skew Macdonald polynomials for ordinary representations
$R=(R,\varnothing)$ are defined as functions of arbitrary time-variables.
Let us define the skew Macdonald polynomials by the expansion
\be\label{Mskew}
M_\lambda\{p^{(1)}+p^{(2)}\}=\sum_\mu M_{\lambda/\mu}\{p^{(1)}\}\cdot M_\mu\{p^{(2)}\}
\ee
which is unambiguous. This implies
\be
M_\lambda\{p^{(1)}+p^{(2)}+p^{(3)}\}=\sum_\mu M_{\lambda/\mu}\{p^{(1)}\}\cdot M_\mu\{p^{(2)}+p^{(3)}\}=
\sum_{\mu,\nu} M_{\lambda/\mu}\{p^{(1)}\}\cdot M_{\mu/\nu}\{p^{(2)}\}\cdot M_\nu\{p^{(3)}\}
\ee
On the other hand,
\be
M_\lambda\{p^{(1)}+p^{(2)}+p^{(3)}\}=\sum_\nu M_{\lambda/\nu}\{p^{(1)}+p^{(2)}\}\cdot M_\nu\{p^{(3)}\}
\ee
Since the Macdonald polynomials form a full basis of symmetric functions at each level, it follows that
\be\label{Macex}
M_{\lambda/\nu}\{p^{(1)}+p^{(2)}\}=\sum_\mu M_{\lambda/\mu}\{p^{(1)}\}\cdot M_{\mu/\nu}\{p^{(2)}\}
\ee

The skew Macdonald can be effectively calculated from the corresponding Littlewood-Richardson coefficients
\be
M_{R/\eta}\{p\} = \sum_\zeta {\bf N}^{R^\vee}_{\eta^\vee\zeta^\vee}\cdot M_\zeta\{p\}
\label{skewviasum}
\ee
where  ${\bf N}$  are the $(q,t)$-dependent structure constants
\be
M_{R_1}\{p\}\cdot M_{R_2}\{p\} = \sum_Q {\bf N}^{Q}_{R_1,R_2} \cdot M_Q\{p\}
\ee
Note that in (\ref{skewviasum}) they depend on transposed diagrams.
One can also define the dual skew-polynomials by the counterpart of formula (\ref{skewviasum})
\be
Q_{R/\eta}\{ p\} = \sum_\zeta {\bf N}^{R}_{\eta\zeta}\cdot Q_\zeta\{p\}
\ee

The Cauchy formula for the skew Macdonald polynomials is
\be\label{Cauchy}
&&\sum_{\xi} (-Q)^{|\xi|} \cdot M_{\xi/\eta_1}\{p\}\cdot \overline{M}_{\xi^\vee/\eta_2}\{p'\}=\\
&=&\exp\left(-\sum_k \frac{Q^kp_kp_k'}{k}\right) \cdot
\sum_\sigma (-Q)^{|\eta_1|+|\eta_2|-|\sigma|}\cdot
\overline{M}_{\eta_1^\vee /\sigma}\{p'\}\cdot M_{\eta_2^\vee /\sigma^\vee}\{p\}
\nonumber
\ee

We will also need the formula that relates the Macdonald polynomials of the conjugated partitions:
\be\label{id}
{h_\lambda\over h_\mu} M_{\lambda/\mu}= \left[{h_{\lambda^\vee}\over h_{\mu^\vee}}
M_{\lambda^\vee/\mu^\vee}\Big(-{q^k-q^{-k}\over t^k-t^{-k}} (-1)^kp_k\Big)\right]_{q\leftrightarrow t}
\ee
It follows from \cite[eq.(B.13)]{AK08}
\be
h_\lambda M_{\lambda}= \left[h_{\lambda^\vee}
M_{\lambda^\vee}\Big(-{q^k-q^{-k}\over t^k-t^{-k}} (-1)^kp_k\Big)\right]_{q\leftrightarrow t}
\ee
and the definition of the skew Macdonald polynomials (\ref{Mskew}).

Similarly,
\be\label{und}
\overline{M}_{\lambda^\vee/\eta^\vee}\Big\{(-1)^{k+1}\overline{p}_k^{(\mu^\vee)}\Big\}=
{h_\lambda \overline{h}_{\eta^\vee}\over \overline{h}_{\lambda^\vee}h_\eta}M_{\lambda/\eta}\Big\{-\underline{p_k}^{(\mu)}\Big\}
={||\eta||^2\over ||\lambda||^2}M_{\lambda/\eta}\Big\{-\underline{p_k}^{(\mu)}\Big\}
\ee
where $\underline{p_k}(q,t):=p_k(q^{-1},t^{-1})$.

\end{document}